\documentclass[journal=jacsat,manuscript=article]{achemso}

\usepackage{chemformula} 
\usepackage[T1]{fontenc} 
\usepackage{todonotes} 
\usepackage{mathptmx}      
\usepackage{mhchem}

\usepackage{subfig}
\usepackage{upgreek} 
\usepackage[range-units=single, uncertainty-mode=separate]{siunitx} 
\usepackage{booktabs}
\usepackage[english]{babel}
\usepackage{xcolor}
\usepackage{amsmath} 
\usepackage{hyperref}

\newcommand*\Zb{\latin{e.g.,}}
\newcommand*\Dh{\latin{i.e.,}}

\author{Philipp Ritzert}
\author{Alexandra Striegel}
\author{Regine von Klitzing}
\email{klitzing@smi.tu-darmstadt.de}
\phone{+49(0)6151-16-24506}
\affiliation[TU Darmstadt]
{Soft Matter at Interfaces, Department of Physics, Technical University Darmstadt, Hochschulstra\ss{}e 8, 64289 Darmstadt, Germany}

\title
  {Ion-specific Stability of Gold Nanoparticle Suspensions}


\begin{document}


\begin{abstract}
\noindent 
Gold nanoparticles (AuNPs) play an important role in fundamental research and development due to their versatile applications and biocompatibility.
This study addresses the aging of three AuNP suspensions after the addition of various sodium salts along the well-known Hofmeister series (NaF, NaCl, NaBr, NaI, NaSCN) at different salt concentrations between \SI{10}{mM} and \SI{100}{mM}.
The AuNP types differ in size (\SI{5}{nm} \latin{vs.} \SI{11}{nm} in diameter) and the capping type (physisorbed citrate \latin{vs.} covalently bound mercaptopropionic acid (MPA)). We monitor the aggregation of the AuNPs and the suspension stability optically (absorption spectroscopy, photography) and by electron microscopy. 
The large range of salt concentrations results in a large variety of colloidal stability, \Dh{} from stable suspensions to fast destabilization followed by sedimentation. At intermediate and high salt concentration strong ion-specific effects emerge that are non-monotonous with respect to the Hofmeister series. In particular, the chaotropic salts, NaI and NaSCN, strongly alter the absorption spectra in very different ways. NaI fuses AuNPs together influencing the primary absorption, while NaSCN retains AuNP structure during aggregation much stronger than the remaining sodium halides, resulting in a secondary absorption peak. 
Although decreasing the size of AuNPs leads to more stable suspensions, the ion specific effects are even more pronounced due to the increase in total available surface. Even the covalently bound MPA capping is not able to stabilize AuNPs against particle fusion by NaI, although it delays the process. Despite the complex interplay between different effects of ions on the stability of colloidal dispersions, this study disentangles the different effects from electrostatic screening, via adsorption at the interface and bridging of AuNPs, to the competition between ions and the capping agent of the AuNPs. These findings are crucial for the fabrication of inorganic/organic composites by targeted assembly of AuNPs in a preexisting matrix controlled by the presence of salt.

\end{abstract}


\section{Introduction}
\label{intro}
Materials that respond to external stimuli have become increasingly popular in recent decades, due to the great advantages they offer in terms of applicability and adaptability to various scientific and everyday problems~\cite{Christau.2015}. They often consist of several components whose interaction can lead to interesting new properties with a high impact in medicine~\cite{Hu.2020,Madkour.2018}, environmental technology~\cite{GalloCordova.2021}, and nanoactuation~\cite{Choi.2008,Zhang.2012}.
 
Gold nanoparticles (AuNPs) are frequently used in smart materials due to the advantages they bring to the composite. AuNPs are easy to synthesize, chemically stable and allow a flexible capping choice~\cite{Amendola.2017}. Moreover, they exhibit a  localized surface plasmon resonance (LSPR) in the visible spectrum that depends on particle specific and environmental parameters. In combination with an organic matrix, they serve as sensors for specific ions~\cite{Liu.2011,Wang.2013}, pH~\cite{Boyaciyan.2018}, temperature~\cite{Lange.2012}, and solvents~\cite{Tong.2022}. The specific application also depends on the matrix but the distribution of the AuNPs in the matrix is crucial for efficient performance. At small AuNP separation (below the particle diameter) plasmon coupling occurs leading to separation-dependent absorption spectra~\cite{Lange.2012}. Sensing these changes allows the determination of interparticle distances so that the AuNPs are referred to as `nano ruler'~\cite{Caselli.2021}. In addition, AuNPs can serve as `hot spots' due to the heat transfer into the environment after plasmon excitation. If the matrix is thermosensitive, illumination near the LSPR may trigger the heat-induced response. This was shown for the laser induced shrinking of thermosensitive microgels with incorporated AuNPs~\cite{Lehmann.2018}.

The recent advances in applications very often expose composite materials to saline surroundings -- either as a result of the core function or concomitantly, \Zb{} in the human body or environmental applications~\cite{Facchi.2017, vanRie.2017}. Effects of salt solutions on AuNP-based materials have been studied for quite some time -- both for fundamental understanding and specific applications~\cite{Kaulen.2018,Li.2021,Wang.2020,Altansukh.2016}. The DLVO theory (after Derjaguin, Landau, Verwey, and Overbeek) describes the electrostatic (de)stabilization of NP suspensions or composite materials considering the overall concentration of present salts and surface charges~\cite{Israelachvili.2011, Mateos.2022,Ludwig.2021}. Furthermore, AuNPs are known to exhibit ion-specific effects regarding aggregation in aqueous suspensions~\cite{Seela.2005}. Composite materials containing AuNPs inherit these ion-specific effects in saline solutions~\cite{Christau.2017}. Besides the utilization for sensing applications, ion-specificity recently also drew interest to exert control over the fabrication process~\cite{Zhang.2014}. The main advantage lies in the distinct ion-specific effects shown by matrix and AuNPs, which enables fine-tuning of the resulting composite materials.

Ion-specific effects usually follow the Hofmeister series that places anions and cations in an order between cosmotropic (water structure-making) and chaotropic (water structure-breaking)~\cite{Gregory.2022, Kunz.2010}. The historic naming refers to an explanation model that associates the ion-specific effects to the order of water molecule around the anion and cation. Nowadays, it's known that ions barely disturb the water structure beyond the first layer around the ion~\cite{Gregory.2022,Omta.2003}. However, the discrimination between well-solvated cosmotropes, which avoid interactions with surfaces in aqueous solution, and the poorly solvated, surface-affine chaotropes is still under debate. The Hofmeister series dates back to the late 19th century~\cite{Hofmeister.1888}, ordering anions with their ability to stabilize protein solutions, and since then proved very reliable for various observables in aqueous systems~\cite{dosSantos.2010, Rana.2022, Wei.2023, Kang.2020}. Similarly, investigations of colloidal stability showed systematic variation along the Hofmeister series~\cite{Schelero.2011,Merk.2014}. Until today, the mechanism behind the Hofmeister series is not fully understood despite numerous studies on various systems including polymers~\cite{Dodoo.2011,Zhang.2005}, complex mixed-salt solutions~\cite{Robertson.2021}, and considering various ion parameters as control variables~\cite{Kunz.2010, Merk.2014, Gregory.2021}. An excerpt of the anion series important for this work reads: $\ch{F^-} < \ch{Cl^-} < \ch{Br^-} < \ch{I^-} < \ch{SCN^-}$, where $A<B$ denotes a higher surface affinity of $B$. An extensive illustration and discussion can be found elsewhere~\cite{Volodkin.2014,Mazzini.2017}. 

Treatment of composite materials with salt solutions can lead to ion-specific, (ir)reversible structural changes~\cite{Christau.2017}. Therefore, it is a powerful potential tool for controlling AuNP assembly in a matrix. Earlier studies already demonstrated ion-specific rearrangement of AuNPs immobilized inside a polymer matrix~\cite{Christau.2017}, where possible AuNP aggregation depended on the applied salt. This leads to ion-dependent reversibility of the AuNP rearrangement, which was tracked by UV-vis absorption spectroscopy. However, efficient utilization of (mixed-) salt stimuli requires detailed knowledge about the ion-specific response of the individual components. Considering the large number of salts and wide range of relevant concentrations the ion-specific stability of AuNP suspensions is still under discussion. 

Ion-specific stability of AuNP suspensions has been investigated for pristine (uncapped) AuNPs showing a systematic stabilization along the direct Hofmeister series (increasing suspension stability from \ch{F-} to \ch{I-})~\cite{Merk.2014}. Here, the salt concentrations remain below \SI{1}{mM}. The absence of a stabilizing capping eliminates many possibly disturbances that impair assessment of the pure ion-gold interaction. However, it also limits the range of salt concentration when considering more complex environments, both from an fundamental and application point-of-view, since uncapped AuNPs will aggregate at relatively low salt concentrations. 

To increase colloidal stability AuNP are often coated with a stabilizing agent. A prominent example is AuNPs coated with poly(ethylene glycol)~\cite{Zhang.2017}. Vaknin and coworkers investigated the formation of two-dimensional AuNP superlattices at the water-vapor interface by varying both the cation and anion of added salt (limiting the anions to the cosmotropic part of the series). They found these superlattices for a variety of mono- and divalent salts and a wide range of salt concentrations (\SI{5}{mM}--\SI{1500}{mM}). The lattice structure originates from the depletion of ions from the capping layer leading to an ion-specific lattice constant. The authors explain the lattice structure with the salt-induced collapse of the poly(ethylene glycol) capping, but interestingly, the formation even occurred at salt conditions where pure poly(ethylene glycol) chains do not show any phase separation. Subsequently, the same group showed that the presence of chaotropic anions \ch{I-} prohibits the formation of a well-defined superlattice\cite{Wang.2020}. They attribute this effect to a significantly stronger shrinking of the capping layer combined with the formation of small AuNP clusters, which was not observed for the other salts.

Kaulen and Simon investigated the influence of pH-value and cation (\SI{50}{mM}--\SI{200}{mM}) on the assembly of positively and negatively charged AuNPs  to metallic substrates from aqueous suspension~\cite{Kaulen.2018} (charges controlled by thiol-based cappings). The adsorption of AuNP showed systematic dependence on the specific cation (decreasing adsorption from \ch{Li+} to \ch{Cs+}). Evaluation of the stability of AuNP suspensions in the presence of the various salts revealed distinct differences between the two cappings: the suspensions of negatively charged AuNPs destabilized at lower salt concentrations and showed non-systematic behavior with regard to the Hofmeister series. This results from the possibility of ion-pairing between the varying cation and the negatively charged capping head group which is not possible for the positively capped AuNPs. Since the pH-value is stabilized by buffer molecules these may additionally impact the ion-specific behavior. 

Zhang \latin{et al.} utilized high concentrations of salt (\SI{180}{mM}) to let citrate-coated AuNPs of various sizes assemble into porous gold structures (referred to as `sponge')~\cite{Zhang.2014}. The AuNP size and the choice of anion governed the resulting structure of the gold sponge changing the pore size and porosity. Importantly, the addition of \ch{I-} lead to a change in structure compared to other halides, as the AuNPs formed larger solid structures. This approach of AuNP assembly leads to fast destabilization and large aggregates for all salts. In contrast, for using salts as trigger for AuNP distribution a better tunability of colloidal stability is desired. This demands for studies at lower concentrations of different salts. Furthermore, the question remains open how the colloidal stability depends on the capping molecule, esp. the strength and type of bond that the capping forms with the gold surface.

Therefore, in the present study, we characterize the ion-specific aging of AuNP suspensions for five monovalent sodium salts along the Hofmeister series: NaF, NaCl, NaBr, NaI, NaSCN for salt concentrations between \SI{10}{mM} and \SI{100}{mM}. We monitor the aging of the salt-containing suspensions by photography and absorption spectroscopy (UV-vis) and visualize possible aggregation by transmission electron microscopy (TEM). To get information about the influence of the AuNP surface, we investigate the impact of AuNP size (diameter \SI{5}{nm}, \SI{11}{nm}) and capping molecule (physisorbed citrate, covalently bound mercaptopropionic acid) on the ion-specific effects.

\section{Materials and methods}
\label{sec:materials}
\subsection{Chemicals}
\ch{HAuCl4}, \ch{Na3citrate}, \ch{NaBH4}, thioctic acid, 3-mercaptopropionic acid, and NaSCN were purchased from Sigma Aldrich (now Merck KGaA, Germany). NaF, NaCl, NaBr, NaI were purchased from Merck KGaA and all samples of a series were diluted from the same stock suspension to ensure comparable salt concentration. Ethanol was purchased from Carl Roth GmbH + Co. KG and miliQ grade de-ionized water ($\rho \geq \SI{18.2}{\mega\ohm\centi\metre}$; miliPore, Merck Germany) was used throughout the project.

\subsection{Gold Nanoparticle Synthesis}
The synthesis recipes are all identical or very similar to the indicated literature and only summarized here briefly. The AuNP types are indexed by their size and capping type, where L and S refer to large and small, respectively; the capping is either citrate (Cit) or mecaptopropionic acid (MPA). We synthesized \mbox{L-Cit}, \mbox{S-Cit}, and \mbox{L-MPA}. 

\textbf{L-Cit.\cite{Christau.2014}} For \SI{100}{mL} of \SI{0.01}{wt\percent} suspension \SI{9.84}{mg} (\SI{0.025}{mmol}) \ch{HAuCl4} is dissolved in \SI{97}{mL} water (from a \SI{1}{wt\percent} stock suspension) and heated to boiling in a conical flask while stirring with \SI{350}{rpm}. When the solution starts boiling \SI{3}{mL} of \SI{38.8}{mM} \ch{Na3citrate} solution is added, and the solution is kept boiling. Within the first \SI{5}{min} of boiling, the solution first loses the yellow color, then changes from grayish over blueish and violet to dark red. Afterwards, the color only slightly brightens up. After the boiling for \SI{20}{min}, the solution is left to cool down to room temperature for at least \SI{2}{h} under continued stirring. 

\textbf{L-MPA.\cite{Boyaciyan.2018b,Christau.2016}} To obtain L-MPA the Cit capping was exchanged for MPA in two-step-process to avoid aggregation. First, the L-Cit suspension is stirred with \SI{250}{rpm} in a conical flask. Per \SI{100}{mL} of suspension, \SI{15}{mg} thioctic acid is dissolved in \SI{2.5}{mL} ethanol and added to the suspension changing the color to dark red. \SI{1}{M} NaOH solution is added in \SI{0.25}{mL} steps until the color is red again (approx. \SI{6}{mL}), followed by stirring the suspension overnight. Subsequently, \SI{8.18}{uL} 3-mercaptopropionic acid are added and suspension stirred overnight again. Lastly, AuNPs are precipitated by addition of ethanol (approx. \SI{2}{mL} ethanol per \SI{1}{mL} suspension) and centrifuge \SI{30}{min} at \SI{10000}{rpm} in a Heraeus Multifuge (ThermoScientific). Afterwards, AuNPs are re-suspended in \SI{100}{mL} water and \SI{5}{drops} of NaOH (\SI{1}{M}) are added. 

\textbf{S-Cit.\cite{Boyaciyan.2018b}} For \SI{100}{mL} of final suspension \SI{9.84}{mg} \ch{HAuCl4} is dissolved in \SI{98}{mL} water (from a \SI{1}{\percent} stock suspension), \SI{0.25}{mL} trisodium citrate solution (\SI{0.1}{M}) is added and stirred with \SI{350}{rpm} in a conical flask. \SI{2}{mL} of \ch{NaBH4} solution (\SI{8}{mg} per \SI{2}{mL}) is added to the flask, immediately changing the color to brownish red. The suspension is stirred for a minimum of \SI{2}{h}. 

Since all AuNP suspensions contain the same amount of gold salt, the AuNP number density varies with the AuNP size. \mbox{S-Cit} are smaller than \mbox{L-Cit} by a factor of 2.25, leading to an increase of AuNP number density by a factor of approximately 11 (cf. Figure~\ref{fig:AuNPOverview} for diameters of each AuNP type). 

Mixing of suspensions always involves preparation of the salt solution (with twice the final concentration) and subsequent addition of the AuNP suspension (of identical volume) to minimize the time lag between samples of the same series. Further, this procedure ensures identical mixing conditions for all samples.

\subsection{Photography} 
Samples are prepared in standard UV-vis cuvettes of PMMA to ensure comparability with UV-vis spectra. No significant evaporation of the suspensions within the three days of measurement could be observed. Photos of one AuNP type are always taken for all concentrations simultaneously (\SI{10}{mM} to \SI{50}{mM}) with one photo every \SI{15}{min}. Only \SI{100}{mM} photos were taken in a separate series. The mixing is done for one concentration at a time to reduced lag times within one concentration to a few seconds. The time between mixing and the first photo was \SI{3}{min}. Preparation was performed from lowest to highest concentration with \SI{5}{min} delay between each concentration, since the aging is much slower at low concentrations. To eliminate apparent color changes due to illumination variation all photos contain a reference suspension without salt and photos of one AuNP type are always taken together. The camera is a Nikon D7200 with macro lens (TAMRON, SP 90 mm, F/2.8).

\subsection{UV-vis Absorption Spectroscopy}
Samples are mixed again \SI{3}{min} before the start of the first measurement in standard PMMA cuvettes. Measurements were performed with a Lambda 650 spectrophotometer (PerkinElmer, Shelton CT USA) scanning from \SI{800}{nm} to \SI{400}{nm} in \SI{1}{nm} steps. Scans were performed with \SI{266}{nm\per s} and \SI{2}{nm} wavelength selection slit width. Prior to a measurement series a calibration of \SI{0}{\percent} and \SI{100}{\percent} transmission was performed. Sample order was kept during all aging experiments to ensure identical sample age.

\subsection{Transmission Electron Microscopy}
Transmission electron microscopy (TEM) is done in the Institute for Material Science (TU Darmstadt) with a FEI CM20 microscope (now Thermo Fisher Scientific, Waltham MA US) with \ch{LaB6} anode. TEM samples are deposited on carbon-covered copper grids (300mesh, Science Service, Germany) for \SI{5}{min} from suspension. For L-MPA suspensions, the drying time was increased to \SI{15}{min}, since fewer L-MPA adsorbed to the grid.

\subsection{Zeta Potential}
Measurements are performed with a Zetasizer Nano ZS (Malvern, UK) in standard DTS1070 PMMA cells with gold electrodes at controlled \SI{25}{\degreeCelsius}. The Zeta potential is calculated using the H\"{u}ckel approximation ($\kappa a < 1$) for salt-free AuNP suspensions using typical parameters for water: viscosity $\eta = \SI{0.8872}{\milli\pascal\second}$, refractive index $n=\num{1.330}$, and dielectric constant $\varepsilon_\mathrm{r} = \num{78.5}$.

%
%
%
%
%
%
%
%

\section{Results}
\label{sec:results}
\begin{figure}
	\includegraphics[width=0.85\textwidth]{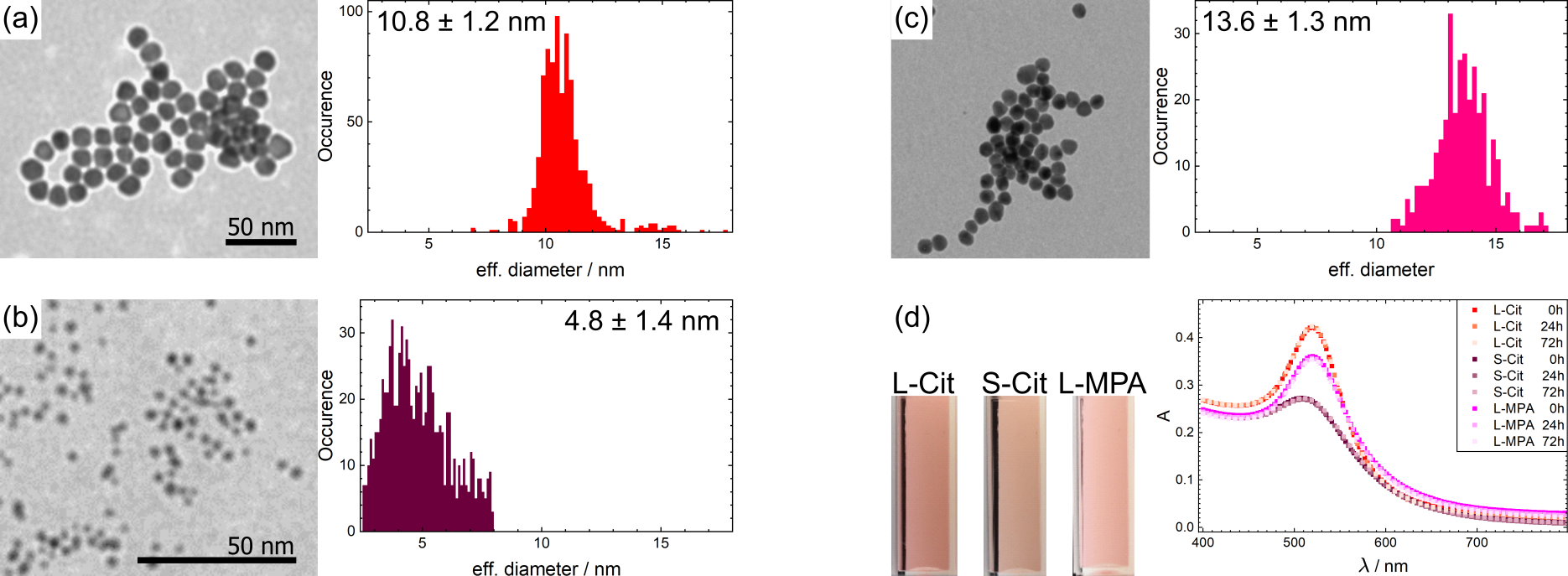}
	\caption{Comparison of basic AuNP parameters for all used AuNP types (for details of labeling see text). Exemplary TEM images for calculation of AuNP diameter and the resulting diameter histograms are summarized for \mbox{L-Cit} (a), \mbox{S-Cit} (b), \mbox{L-MPA} (c). Figure (d) illustrates the resulting optical properties as photos of the salt-free AuNP suspensions and corresponding absorption spectra in the visible range over \SI{72}{h}.}
	\label{fig:AuNPOverview}
\end{figure}
The present work compares two sizes of AuNPs (labeled small (S) and large (L)) stabilized with one of two capping molecules: physisorbed citrate anions (Cit) and covalently bound 3-mercaptopropionic acid (MPA). Overall we consider three AuNP systems: large and small AuNPs with citrate capping and large AuNPs with MPA capping. The capping molecules can be de-protonated providing both electrostatic inter-particle repulsion and hydrogen bonding.  
Figure~\ref{fig:AuNPOverview} shows exemplary TEM images and the respective AuNP diameter histograms taken over multiple TEM images ($> 800$ NPs for \mbox{L-Cit}/\mbox{S-Cit}, $>300$ NPs for \mbox{L-MPA}). The arithmetic mean of the diameter distributions for \mbox{L-Cit}, \mbox{S-Cit} and \mbox{L-MPA} yields \SI{10.8(12)}{nm}, \SI{4.8(14)}{nm}, and \SI{13.6(13)}{nm}, respectively.  

\begin{figure}
	\includegraphics[width=0.45\textwidth]{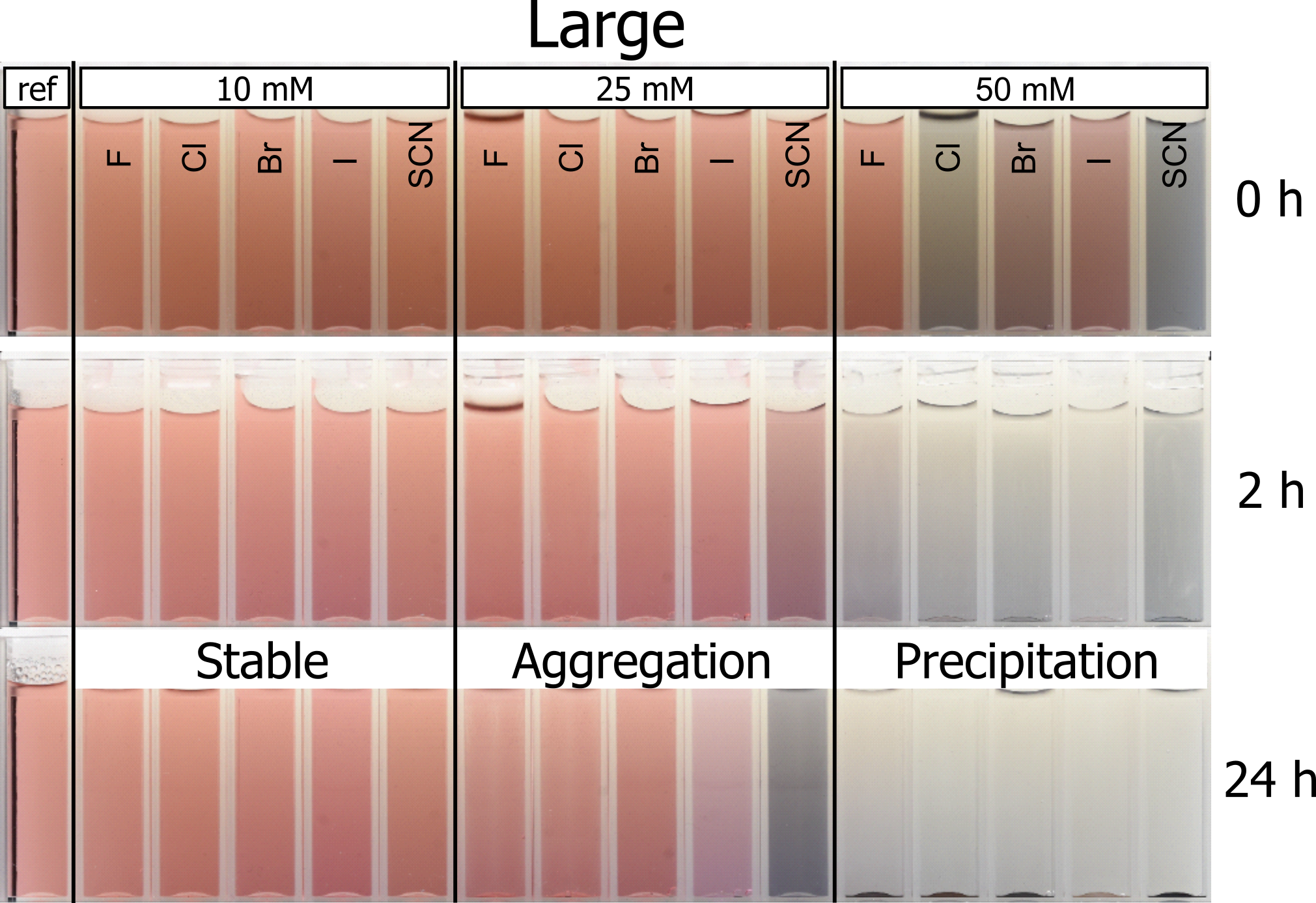}
	\caption{Photo series of \mbox{L-Cit} suspensions mixed with various sodium salts varying along the Hofmeister series: NaF, NaCl, NaBr, NaI, NaSCN. Three values of the salt concentration are separated by the main columns of the graph. The sample on the left-hand side is a salt-free suspension of identical AuNP concentration for reference of the color change. The three rows correspond to different times of measurement after adding salt to the AuNP suspensions. }
	\label{fig:photos_LCit}
\end{figure}
Absorption spectra show distinct absorption peaks (\mbox{L-Cit}: \SI{516}{nm}, \mbox{S-Cit}: \SI{508}{nm}, \mbox{L-MPA}: \SI{520}{nm}). The spectrum of L-MPA is slightly shifted to higher wavelengths and the peak is a bit broader compared to the respective L-Cit spectrum. The suspensions are reddish colored and transparent. Color differences resulting from the varying absorption peaks are visible to the naked eye (Figure~\ref{fig:AuNPOverview}~(d)). Suspensions of \mbox{S-Cit} appear slightly more brownish due to the blue-shifted peak position and the wider absorption peak. Absorption spectra do not indicate changes of the salt-free suspensions for at least three days, which is the investigated time frame in this work. Salt-free suspensions are even stable for months after synthesis, concluded from the unchanged absorption spectra, proving sufficient stabilization of the AuNPs. Zeta Potential values of AuNPs are (in order of the photos in Figure~\ref{fig:AuNPOverview}, (a)): \mbox{L-Cit}: \SI{-58.1}{mV}, \mbox{S-Cit}: \SI{-38.3}{mV}, \mbox{L-MPA}: \SI{-52.7}{mV}. The zeta potential increases with size but does not vary much with capping agent.

\subsection{Suspensions of Large Citrate-Capped AuNPs (L-Cit)}
\label{sec:LCit}
\begin{figure}
	\includegraphics[width=0.85\textwidth]{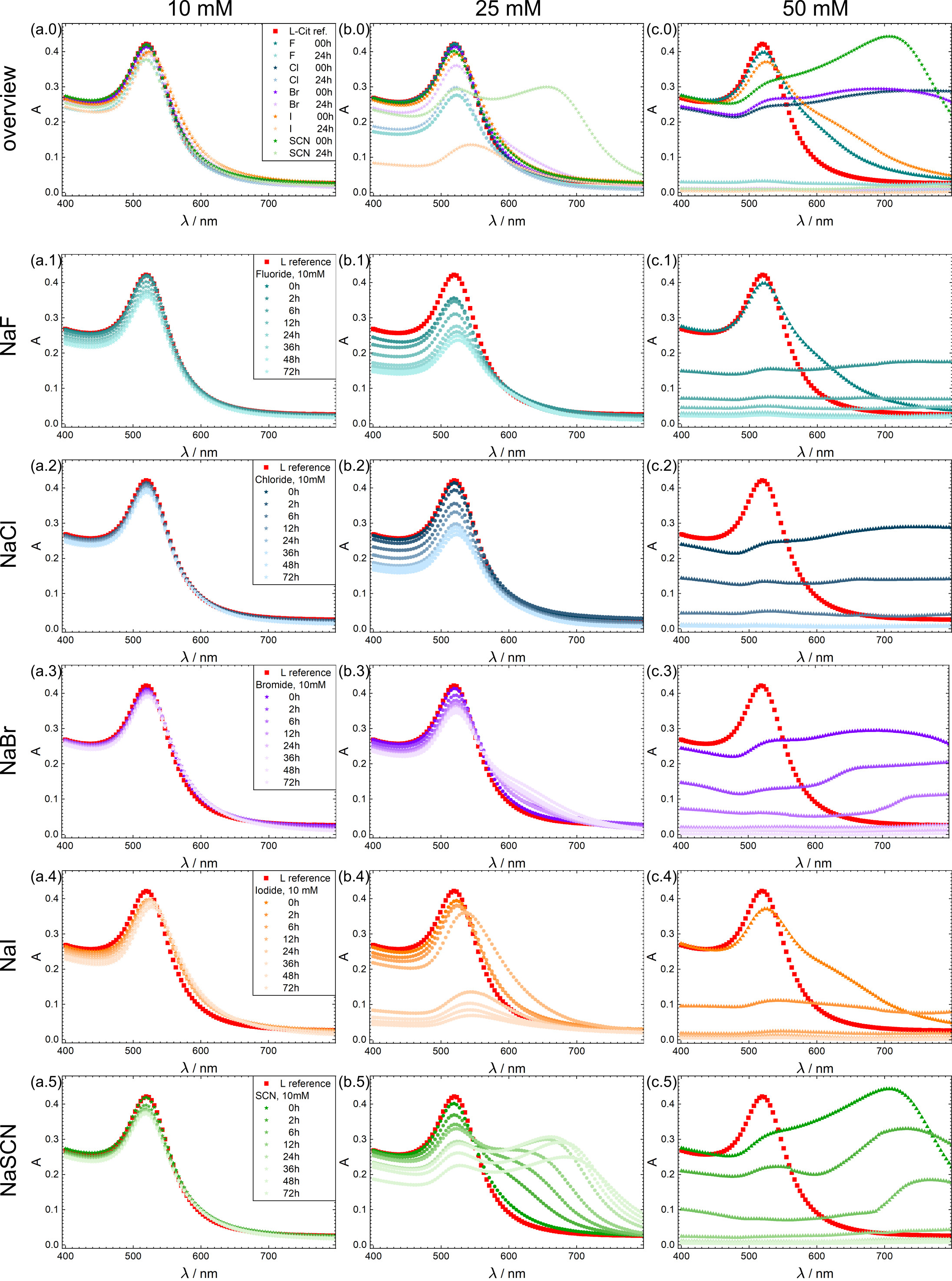}
	\caption{Absorption spectra of \mbox{L-Cit} suspensions mixed with five sodium salts (anions: F, Cl, Br, I, SCN) monitored over time. The concentration of the salt increases with \SI{10}{mM}, \SI{25}{mM}, and \SI{50}{mM} from left to right. The zero row (a.0-c.0) summarizes the spectra of all salts immediately after mixing (strong colors) and \SI{24}{h} later (light colors). The different anions are separated by color type. The remaining rows display the aging of individual anions where lighter colors encode longer aging times.}
	\label{fig:timeMatrix_LCit}
\end{figure} 
Figure~\ref{fig:photos_LCit} presents snapshots of \mbox{L-Cit} suspensions mixed with sodium salts (NaF, NaCl, NaBr, NaI, NaSCN) of three different concentrations (\SI{10}{mM}, \SI{25}{mM}, \SI{50}{mM}) at multiple aging times (\SI{0}{h}, \SI{2}{h}, \SI{24}{h}). The snapshots include a salt-free reference suspension (cuvette on the left). For lowest salt concentration (\SI{10}{mM}) the suspensions show barely any significant color shift. The intermediate salt concentration regime (\SI{25}{mM}) shows no effects after mixing and only minor color changes after \SI{2}{h} for iodide and thiocyanate. These effects grow more prominent after \SI{24}{h}, showing strong color changes for SCN and considerable decrease in color strength for iodide. At \SI{50}{mM} salt concentration, the suspensions change color immediately after mixing ($<\SI{5}{min}$) and the changes strongly vary with the type of anion. After \SI{2}{h} the suspensions are almost completely colorless. The photo series including additional aging times can be found in Figure~\ref{si:photos_LCit} in the SI.

UV-Vis spectroscopy was used to give more details about the effect of different salts and concentrations. Figures \ref{fig:timeMatrix_LCit} (a.0--c.0) compare absorption spectra of the suspensions immediately after mixing and \SI{24}{h} later, separated by the salt concentration in columns. At \SI{10}{mM} salt concentration, the spectra reveal minor changes for iodide and thiocyanate that were barely detectable by eye. At \SI{25}{mM} salt concentration, the intensity of the primary absorption peak decreases for all anions except bromide. The addition of NaSCN generates a secondary absorption peak at around \SI{670}{nm} related to the light-blue color in the photos. Iodide induces a much stronger decrease of the primary absorption paired with a \mbox{red-shift} and broadening of the absorption peak. At \SI{50}{mM} salt concentration, the spectra after mixing reflect the wide variety of colors in Figure~\ref{fig:photos_LCit}. For thiocyanate the secondary absorption peak appears even stronger, while iodide already shows a slight \mbox{red-shift} of the absorption peak. Chloride and bromide cause similar patterns with almost constant absorption over the whole visible spectrum. For fluoride and iodide, only a minor absorption shoulder appear at higher wavelengths. 

To get deeper insights into the aging, Figures \ref{fig:timeMatrix_LCit} (a.1--c.5) compare the absorption spectra of the \mbox{L-Cit} suspension in the presence of salt over time up to \SI{72}{h} after mixing. The color of the plot symbols encodes the aging time between salt addition and the measurement, changing from strong to light colors with increasing aging time. Following the \SI{10}{mM}-column (a.1--a.5), the salt concentration is too low to lead to major salt effects even for aging times up to \SI{72}{h}. Along the \SI{25}{mM}-column (b.1--b.5), the primary absorption peak decreases with time for all salts. From fluoride to bromide, the absorption reduction becomes less pronounced. For \ch{Br-} a slight absorption shoulder is detectable.  

\begin{figure}
	\includegraphics[width=0.30\textwidth]{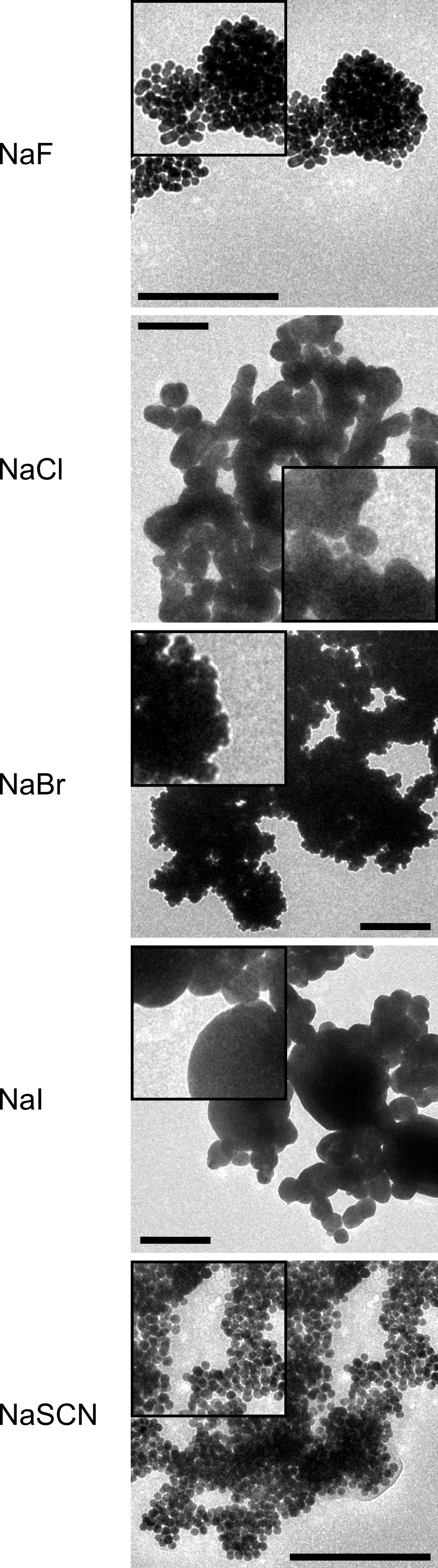}
	\caption{TEM images of aggregates formed in \mbox{L-Cit} suspensions for \SI{6}{h} after addition of \SI{50}{mM} of the respective salt. Different anions ordered vertically along the Hofmeister series as before: \ch{F-}, \ch{Cl-}, \ch{Br-}, \ch{I-}, \ch{SCN-}. The scale bars in all images are \SI{200}{nm}. To facilitate size comparison of the aggregates framed insets represent a \SI{200}{nm}-cutout. Aggregation took place in suspension and aggregates assembled on a TEM grid for only \SI{5}{min}.}
	\label{fig:TEMseries_LCit}
\end{figure}
Aging behavior drastically changes for iodide as the decrease of the primary absorption peak at \SI{25}{mM} is relatively minor at early stages, but the absorption almost completely vanishes between \SI{12}{h} and \SI{24}{h}, indicating precipitation. This trend is even more pronounced at \SI{50}{mM} (c.1--c.5): all salts induce precipitation but point in time and manner differ strongly. Iodide only induces the formation of a prominent shoulder at high wavelengths directly after mixing while the other sodium halides broaden the absorption peak into a plateau over the whole visible spectrum. However, iodide induces the fastest precipitation of all salts.

Interestingly, the resulting effects of thiocyanate addition do not continue the trend of iodide. At \SI{25}{mM}, thiocyanate induces a reduction of the primary absorption peak while not changing its position. Moreover, a secondary absorption peak forms: first visible as a shoulder, then growing between \SI{12}{h} and \SI{24}{h}, and  shifting to higher wavelengths. This secondary absorption peak also emerges for \SI{50}{mM} thiocyanate but the formation and the fading due to suspension instability takes place in less than \SI{12}{h}. 

Figure~\ref{fig:TEMseries_LCit} summarizes the TEM images of \mbox{L-Cit} after \SI{6}{h} aging to illustrate the nanoscopic structure of AuNP aggregates. Additional, shorter aging times (\SI{10}{min}, \SI{2}{h}) are compiled in Figure~\ref{si:TEMseries_LCit} in the SI, resolving the aging process further. The TEM images show a strong aggregation for all salts, which agrees with the results from the absorption spectra. Interestingly, the structure of the aggregates discriminates two substructures: I) conventional aggregation of the AuNPs while retaining their particle-like substructure; II) large, solid, roughly spherical particles with very wide size distribution where the majority of particles are much larger than the initial AuNP diameter of about \SI{11}{\nano\meter}. NaF, NaBr, and NaSCN exhibit substructure I where the individual AuNPs are clearly distinguishable, even though for NaBr this is mainly at the edge of the aggregates. NaI shows clearly substructure II where no AuNP of initial size is present (even after \SI{2}{h}). For NaCl an interesting state in between emerges where the resulting objects are much smaller than for NaI and seem to roughly remain at their size after \SI{2}{h}. Further, the larger objects appear more elongated compared to NaI and a few AuNPs of initial size are still visible.

\subsection{Suspensions of Small Citrate-Capped AuNPs (S-Cit)}
\label{sec:SCit}
\begin{figure}
	\includegraphics[width=0.45\textwidth]{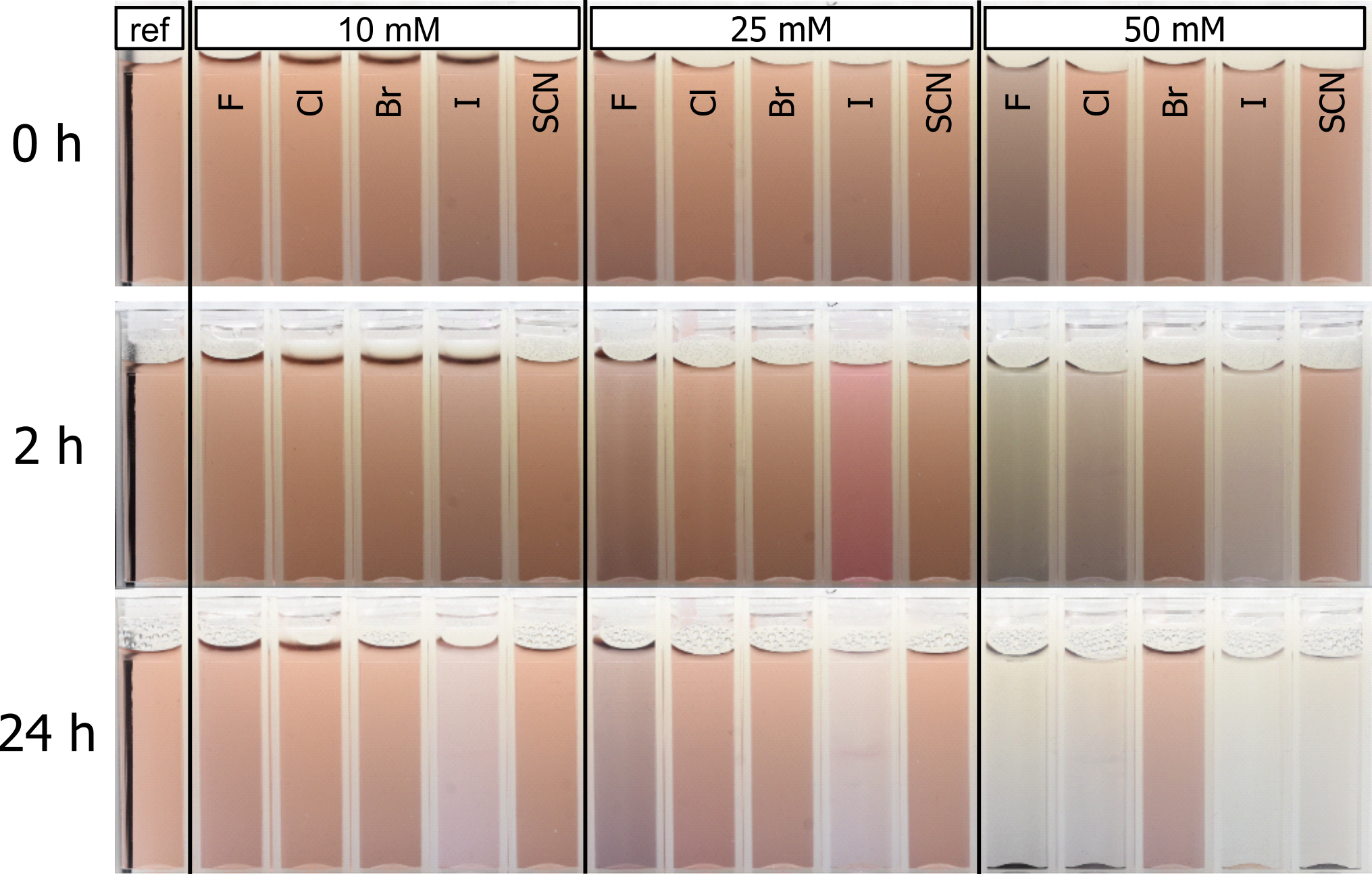}
	\caption{Photo series of \mbox{S-Cit} suspensions mixed with various sodium salts varying along the Hofmeister series: NaF, NaCl, NaBr, NaI, NaSCN. The salt concentration varies from left to right with \SI{10}{mM}, \SI{25}{mM}, \SI{50}{mM}. The far-left sample is a salt-free suspension for reference of the color change. The three rows correspond to different times of measurement after adding salt to the AuNP suspensions.}
	\label{fig:photos_SCit}
\end{figure}
Figure~\ref{fig:photos_SCit} shows a photo series of \mbox{S-Cit} suspensions mixed with salt solutions (NaF, NaCl, NaBr, NaI, NaSCN), analogue to \mbox{L-Cit} suspensions (Figure~\ref{fig:photos_LCit}). The suspensions of \mbox{S-Cit} are more stable that the ones of \mbox{L-Cit}. However, compared to \mbox{L-Cit} ion-specific variations occur equally or even more pronounced. Most prominent, NaI induces color changes at even lower salt concentrations starting at \SI{10}{mM}. The stability increase from NaF to NaBr is more pronounced, even resulting in a stable suspension with \SI{50}{mM} NaBr. To induce destabilization, \mbox{S-Cit} suspensions were studied at \SI{100}{mM} salt concentration (Figure~\ref{si:photos_SCit} in the Supporting Information (SI)). A full destabilization of the suspension happens very quickly for all salts. 

\begin{figure}
	\includegraphics[width=0.85\textwidth]{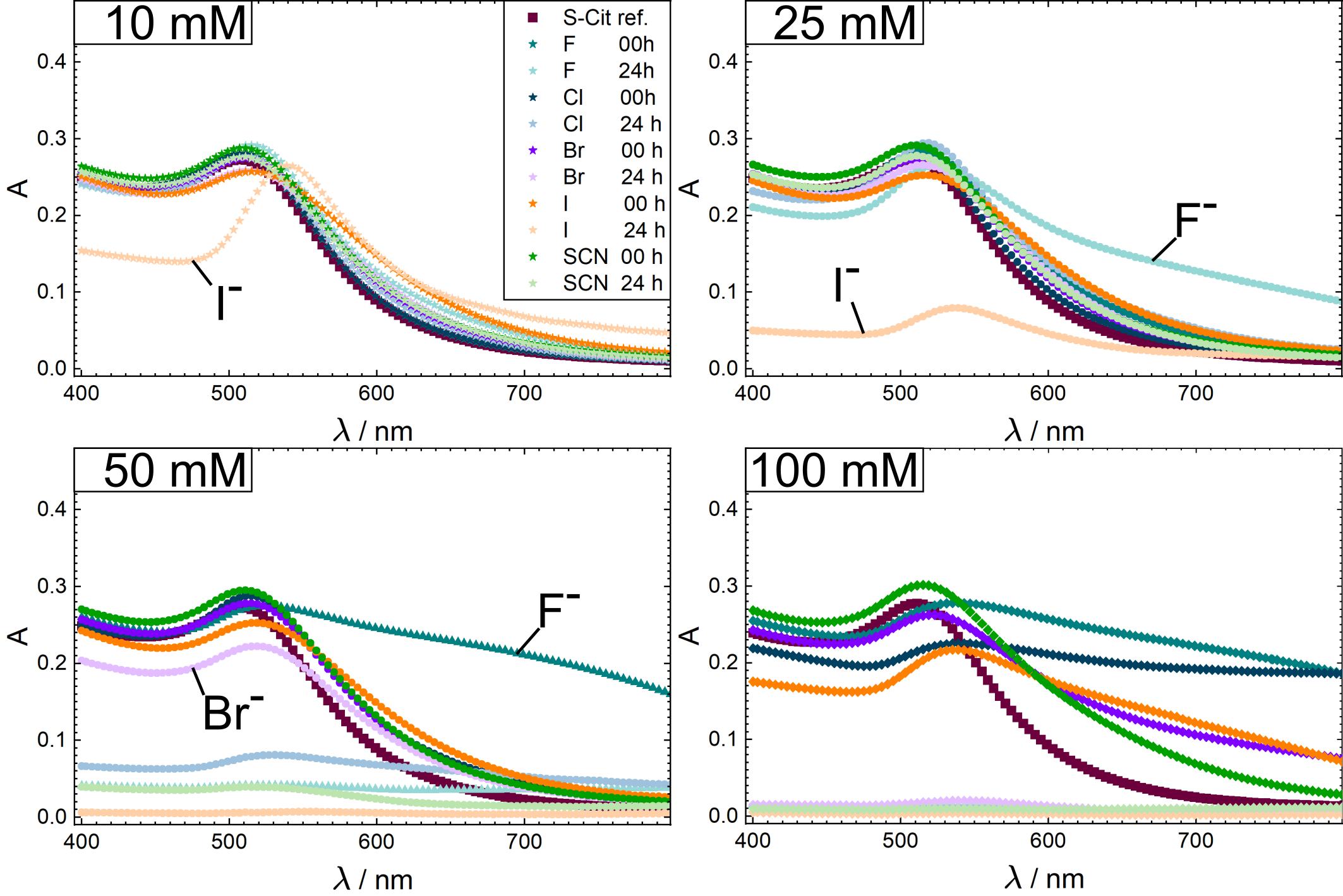}
	\caption{Absorption spectra of \mbox{S-Cit} suspensions mixed with various sodium salt solutions (coded by color) of various concentration immediately after mixing (strong colors) and 24 h later (light colors). The total salt concentration varies along the graphs: \SI{10}{mM}, \SI{25}{mM}, \SI{50}{mM}, \SI{100}{mM}. }
	\label{fig:spectra_SCit}
\end{figure}
Figure~\ref{fig:spectra_SCit} shows a comparison of the absorption spectra of \mbox{S-Cit} in the presence of the sodium salts with the concentration varying from \SI{10}{mM} to \SI{100}{mM} directly after mixing and \SI{24}{h} later. Overall, the spectra agree with the observation of the photos in Figure~\ref{fig:photos_SCit}: minor response of most salts at salt concentrations up to \SI{50}{mM}, and a fast change to broad, featureless absorption at \SI{100}{mM} followed by complete precipitation. For \SI{25}{mM} and \SI{50}{mM} NaF, a pronounced response occurs in the form of a broad shoulder in the high wavelength region, bridging the way to the plateau behavior observed at \SI{100}{mM}. The most striking feature results from NaI at \SI{10}{mM}, where the primary absorption peak shifts to higher wavelengths (similar to \mbox{L-Cit}), but simultaneously the peak also becomes more distinct, \Dh{}, the relative width decreases. As for \mbox{L-Cit}, NaI induces the most significant destabilization (\mbox{S-Cit}: \SI{25}{mM}). Simultaneously, the corresponding effect of NaSCN on \mbox{L-Cit} (formation of a secondary absorption at high wavelength) is completely missing. Instead, suspensions containing NaSCN behave similar to NaCl for all concentrations. 
\begin{figure}
	\includegraphics[width=0.85\textwidth]{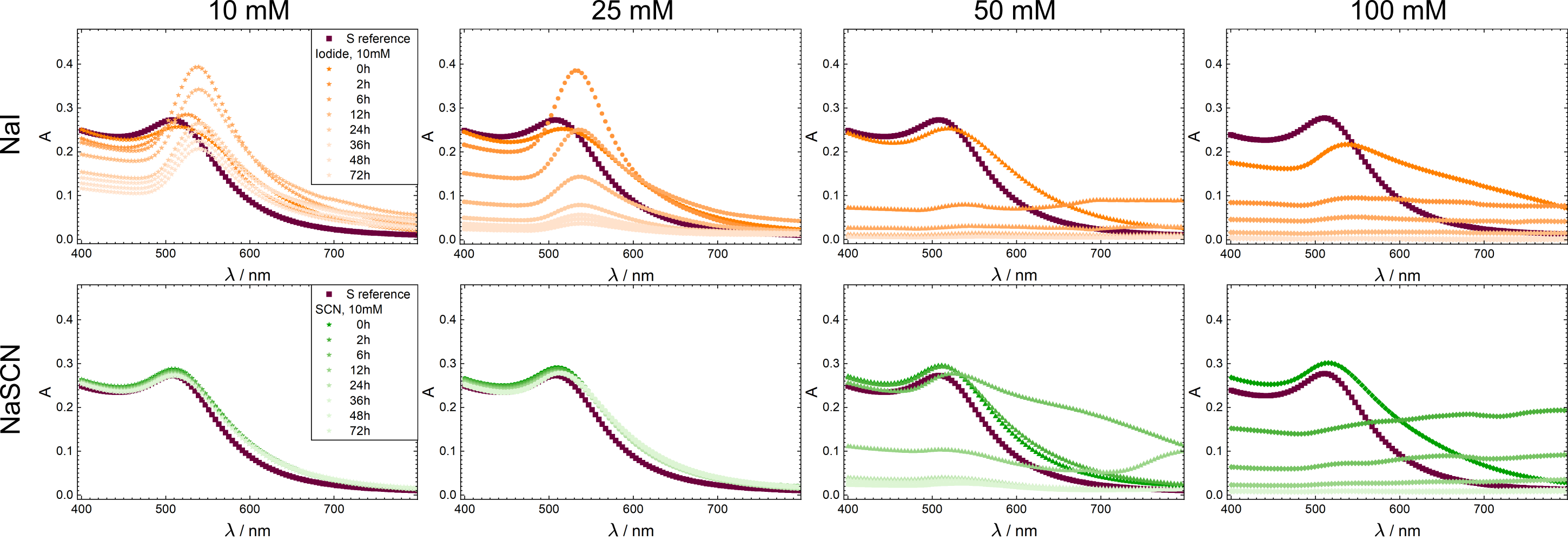}
	\caption{Time series of \mbox{S-Cit} suspensions mixed with NaI and NaSCN of various concentration ranging from \SI{10}{mM} to \SI{100}{mM}. The color scheme agrees with earlier Figures regarding color type and strength. }
	\label{fig:timeMatrix_SCit}
\end{figure}

A detailed representation of the aging of \mbox{S-Cit} suspensions containing various amounts of NaI and NaSCN is shown in Figure~\ref{fig:timeMatrix_SCit}. For iodide, the time series shows the continuous \mbox{red-shift} of the primary absorption peak with simultaneous increase in peak intensity and decrease of relative peak width. This process is almost identical for \SI{10}{mM} and \SI{25}{mM}, yet slightly faster for higher salt concentration. At \SI{100}{mM} salt concentration, AuNPs aggregate and precipitate immediately after mixing. The thiocyanate response indeed seems to be very weak compared to the \mbox{L-Cit} case, but at \SI{50}{mM} a strong shoulder at higher wavelengths occurs that indicates secondary peak, yet less distinct than for \mbox{L-Cit}. Aging series' for all salts are compared in Figure~\ref{si:timeMatrix_SCit} in the SI. 

\begin{figure}
	\includegraphics[width=0.85\textwidth]{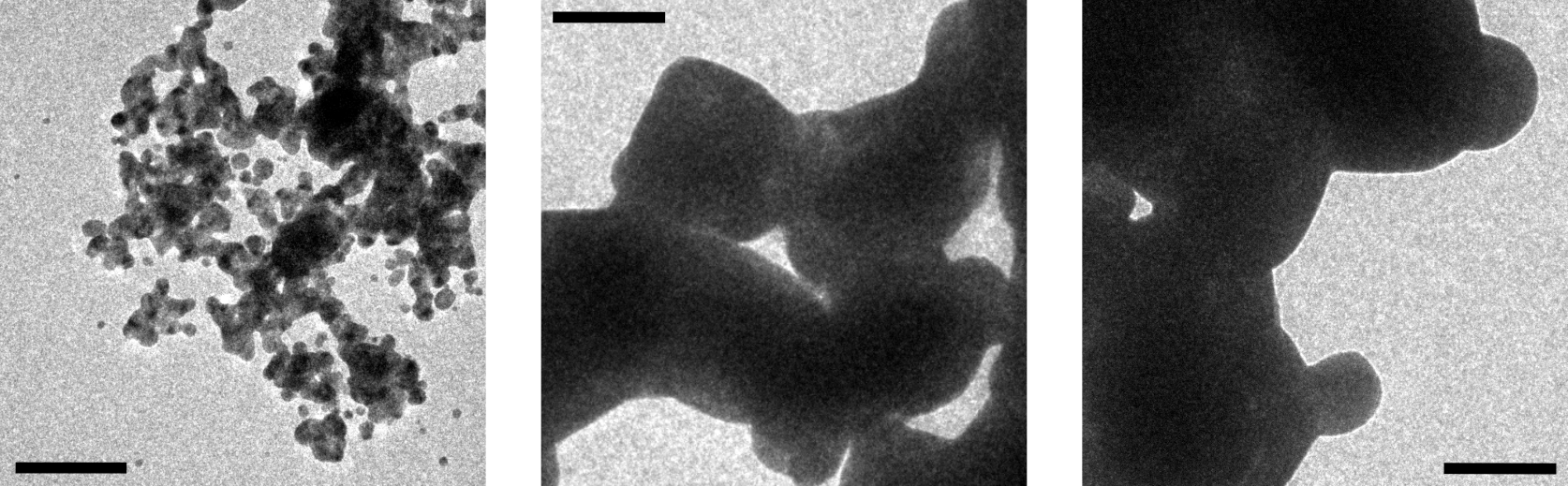}
	\caption{Aging time series (\SI{10}{min}, \SI{2}{h}, \SI{6}{h}) of \mbox{S-Cit} suspensions containing \SI{50}{mM} NaI. The scale bars in all images correspond to \SI{100}{nm}.}
	\label{fig:TEMseries_SCit}
\end{figure}

Figure \ref{fig:TEMseries_SCit} displays TEM images of the aging of \mbox{S-Cit} suspensions containing \SI{50}{mM} NaI. The images clearly illustrate the fast aggregation observed in the absorption spectra. After \SI{10}{min} the AuNPs already mainly lost their spherical shape and are in the process of fusion. The process is mainly finalized after \SI{2}{h}, and afterwards only infrequent growth of the fused structures is observed. The aggregation in presence of NaI is initially faster for \mbox{S-Cit} compared to the \mbox{L-Cit} discussed above.

\subsection{Suspensions of Large MPA-Capped AuNPs (L-MPA)}
\label{sec:LMPA}
\begin{figure}
	\includegraphics[width=0.45\textwidth]{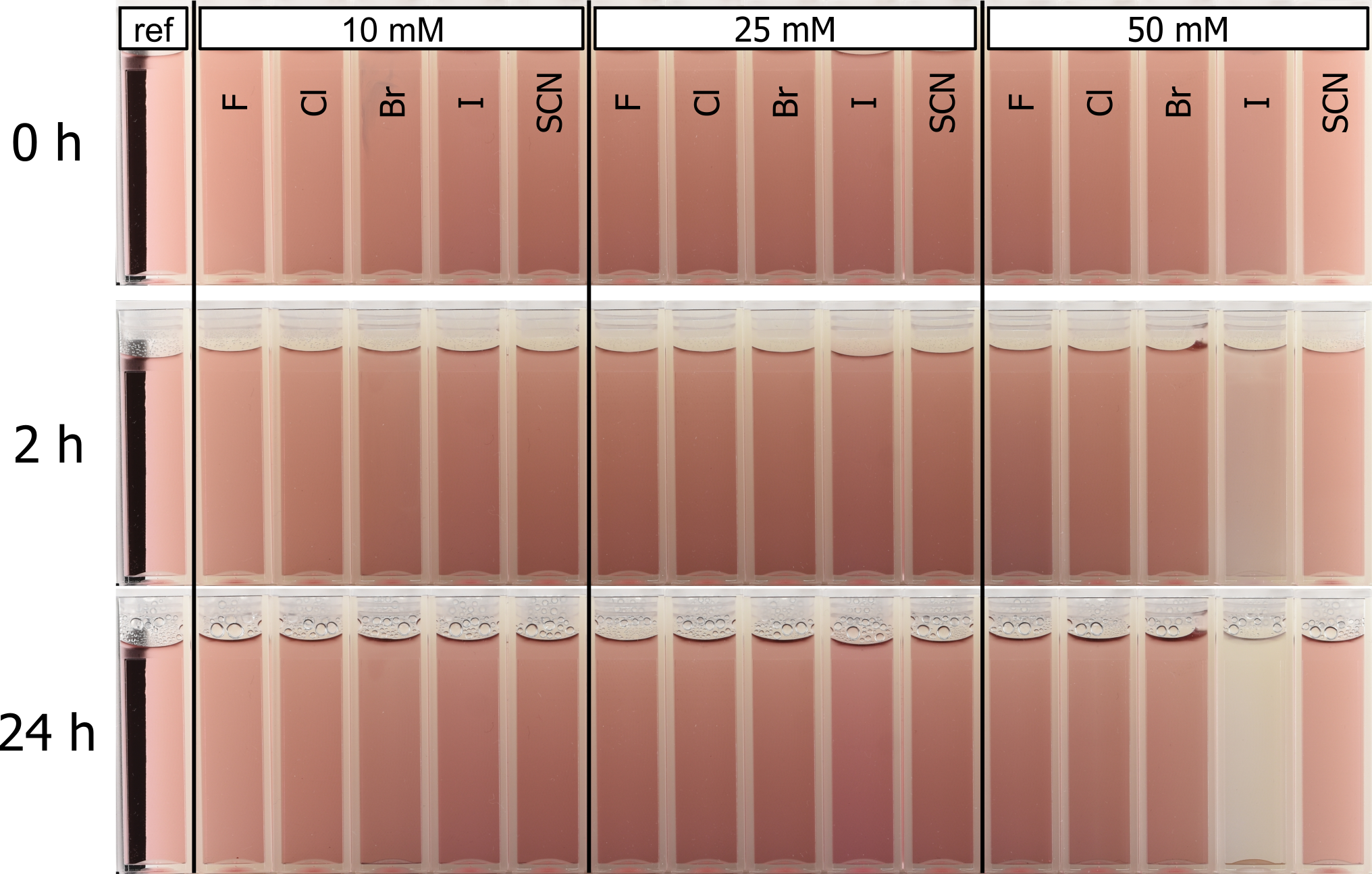}
	\caption{Photo series of \mbox{L-MPA} suspensions mixed with sodium salts, with anions varying along the Hofmeister series: NaF, NaCl, NaBr, NaI, NaSCN. The salt concentration varies from \SI{10}{mM} to \SI{50}{mM} from left to right. The left-most sample is a salt-free suspension for reference of the color change. The three rows correspond to different times of measurement after adding salt to the AuNP suspensions.}
	\label{fig:photos_LMPA}
\end{figure}
Figure~\ref{fig:photos_LMPA} summarizes the photo series of \mbox{L-MPA} as shown before for other AuNP types. Photos at additional aging times and photos for \SI{100}{mM} salt concentration can be found in Figure~\ref{si:photos_LMPA} in the SI. In comparison to \mbox{L-Cit}, the \mbox{L-MPA} suspensions show minimal response at any salt concentration, indicating high suspension stability. The only exception is the addition of NaI at \SI{50}{mM} salt concentration. For \SI{50}{mM} NaI, the suspensions are destabilized very fast, resulting in essentially complete precipitation after \SI{2}{h} while not displaying major response directly after mixing.  

\begin{figure}
	\includegraphics[width=0.85\textwidth]{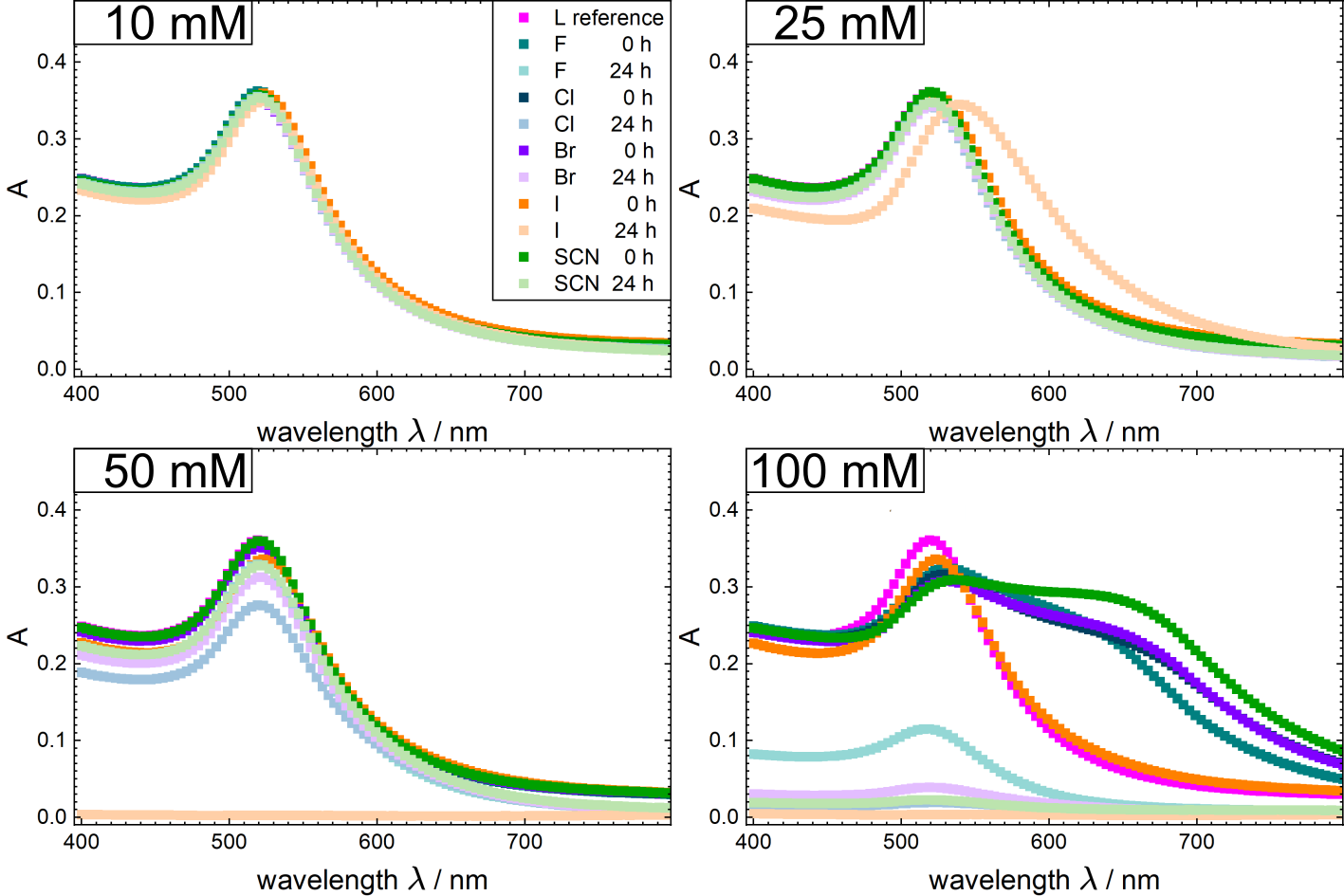}
	\caption{Absorption spectra of \mbox{L-MPA} suspensions mixed with various sodium salt solutions (coded by color type) of various concentration immediately after mixing (strong colors) and \SI{24}{h} later (light colors). The total salt concentration varies along the graphs: \SI{10}{mM}, \SI{25}{mM}, \SI{50}{mM}, \SI{100}{mM}.}
	\label{fig:spectra_LMPA}
\end{figure}
The changes in color from the photo series are resembled in the absorption spectra. The overview absorption spectra for all salt concentrations comparing effects short after mixing and \SI{24}{h} later are shown in Figure~\ref{fig:spectra_LMPA}. Up to \SI{50}{mM}, these spectra show only minor changes in peak position or intensity for any salt except NaI. At \SI{100}{mM}, the addition of either of the other four salts leads to a formation of a relatively non-distinct secondary absorption peak with only minor changes of the primary absorption. After \SI{24}{h}, the suspensions are destabilized and AuNPs precipitated; the remaining absorption of the fluoride \SI{100}{mM} originates from AuNP adsorption to the cuvette walls.

\begin{figure}
	\includegraphics[width=0.85\textwidth]{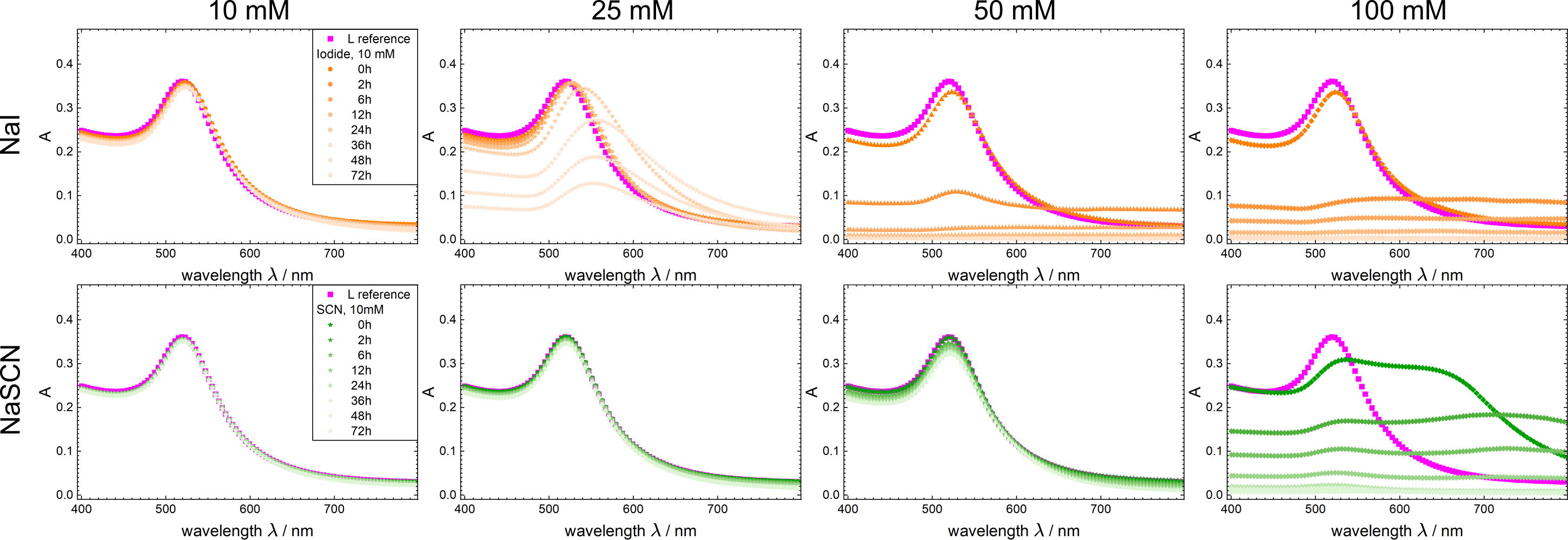}
	\caption{Time series of \mbox{L-MPA} suspensions mixed with NaI and NaSCN of various concentration ranging from \SI{10}{mM} to \SI{100}{mM}. The color code is adapted from respective figures above.}
	\label{fig:timeMatrix_LMPA}
\end{figure}
For NaI and NaSCN aging is displayed in Figure~\ref{fig:timeMatrix_LMPA} for all salt concentrations. Comparing NaI spectra at \SI{10}{mM} and above \SI{50}{mM} salt concentration reproduces the stability-to-destabilization transition observed before, with little immediate response at high salinity but fast precipitation after a minor \mbox{red-shift} of the peak. At \SI{25}{mM}, the suspension once again shows the gradual \mbox{red-shift} accompanied by peak broadening resulting in precipitation over \SI{72}{h}, consistent with the observation of \mbox{L-Cit}. The suspensions with the other salts produce very similar spectra. Therefore, only NaSCN is shown here and the complete aging for all salts and concentrations is summarized in the SI for comparison (Figure~\ref{si:timeMatrix_LMPA}). The suspensions are stable up to \SI{50}{mM} and are destabilized quickly at \SI{100}{mM} salt concentration. Immediately after mixing AuNP suspension and salt solutions, a broad secondary absorption peak appears that flattens into an almost constant plateau during AuNP precipitation. 

\begin{figure}
	\includegraphics[width=0.75\textwidth]{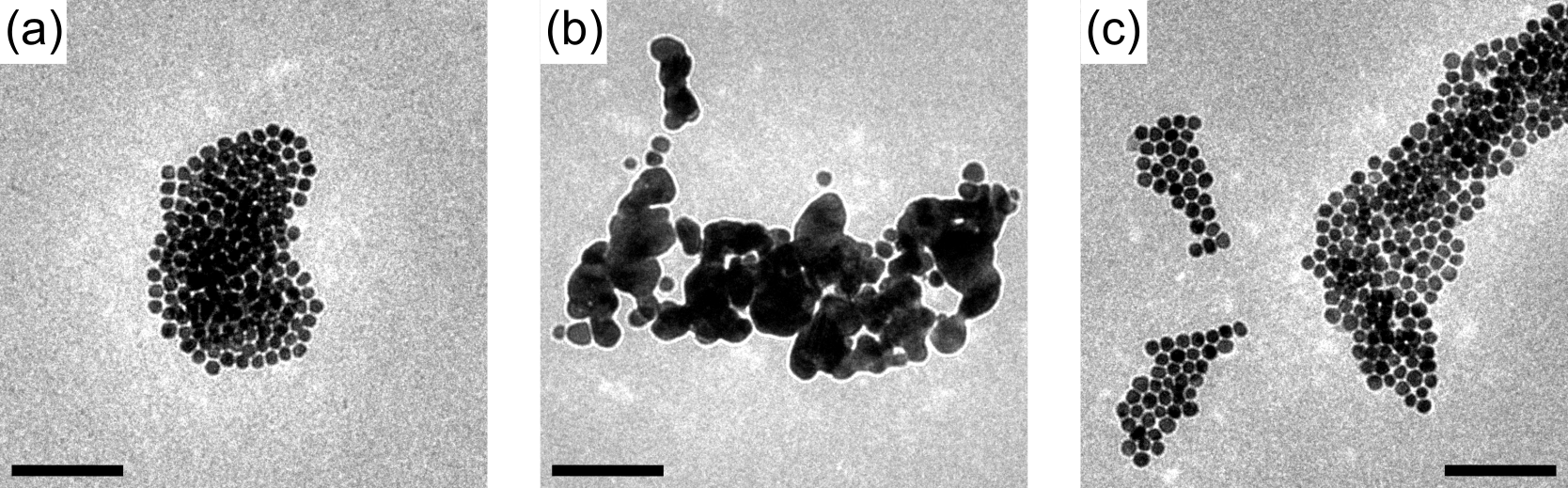}
	\caption{Aging time series of \mbox{L-MPA} after \SI{6}{h} containing selected salts of \SI{50}{mM} concentration: (a) NaBr, (b) NaI, (c) NaSCN. The scale bars in all images represent \SI{100}{nm}. Aggregation took place in suspension and aggregates assembled on a TEM grid for \SI{15}{min}. TEM images of for all salts can be found in the SI (Figure~\ref{si:TEM_LMPA})}
	\label{fig:TEMseries_MPA}
\end{figure}
To illustrate the nanoscopic structure of the \mbox{L-MPA} aggregates, Figure~\ref{fig:TEMseries_MPA} shows TEM images of the aging of \mbox{L-MPA} suspension after \SI{6}{h}, containing \SI{50}{mM} of NaBr, NaI or NaSCN, respectively (all five sodium salts plus a salt-free reference can be found in~\ref{si:TEM_LMPA}). Overall, for all samples (except NaI) no AuNP fusion is observed and the AuNPs adhere predominantly in monolayers as the aggregation is inhibited at \SI{50}{mM} salt concentration. In the case of NaI, \mbox{L-MPA} fuse together, but the resulting structures after \SI{6}{h} are much smaller compared to \mbox{L-Cit} (cf. Figure \ref{fig:TEMseries_LCit}).

\section{Discussion}
\label{sec:discussion}
\subsection{Surface Plasmon Resonance}
The great advantage of studying AuNP suspensions lies in the sensitive response of the displayed localized surface plasmon resonance (LSPR) to changes in the vicinity, for instance changes in the solvent, adsorption of material, or AuNP separation. Photos in Figure~\ref{fig:AuNPOverview} show the red color of the AuNP suspensions that varies slightly between the AuNP types due to size and capping. A detailed discussion of the observed changes in absorption behavior of the salted AuNP suspensions requires the basics of the LSPR origin. These are summarized in the following paragraphs before discussing the influence of salt concentration, specific ion and AuNP type on the prior described results. 

The LSPR of AuNPs in this size range can be well described in dipolar approximation assuming a conducting AuNP-dipole in a surrounding dielectric medium~\cite{Amendola.2017,Trugler.2016}. Respecting monotony of the electrical fields, classic electrodynamics calculations yield a formula for the polarizability $\alpha$ of a AuNP, as a function of the AuNP radius $R$ and optical properties of the AuNP (dielectric function $\varepsilon (\lambda)$) and the surrounding medium ($\varepsilon_\mathrm{m}$)~\cite{Trugler.2016}
\begin{equation}\label{eq:SPRPeakPos}
\alpha = 4\pi R^3 \frac{\varepsilon (\lambda) - \varepsilon_\mathrm{m}}{\varepsilon (\lambda) +2 \varepsilon_\mathrm{m}}\,.
\end{equation}
The polarizability is a complex-valued function and provides the cross-sections of absorption $C_\mathrm{abs} \propto \mathrm{Im}[\alpha]$ and scattering $C_\mathrm{sca} \propto \vert \alpha^2\vert$. Both sum up to the total extinction cross-section~\cite{Trugler.2016}. The dependence of these cross-sections on the AuNP size results in domination of absorption over scattering for the AuNPs below \SI{50}{nm} in diameter. 

In that region the extinction maximum \mbox{red-shifts} with the AuNP size~\cite{Kimling.2006}, which enters only implicitly via the size dependence of $\varepsilon(\lambda)$~\cite{Kreibig.1995}. This dependence directly leads to a blue-shift of the absorption maximum of the \mbox{S-Cit} relative to the \mbox{L-Cit}. In addition, the absorption peak is broader compared to the L-Cit, which is a general feature of LSPR of very small AuNPs~\cite{Amendola.2017,Bonaccorso.2013}. This effect occurs even more pronounced here due to the wider relative size distribution of the \mbox{S-Cit}. Combining the differences in peak position and peak width lead to the slightly more brownish color of the \mbox{S-Cit} suspension compared to the \mbox{L-Cit}. 

In addition, Eq. (\ref{eq:SPRPeakPos}) yields the increase of strength of absorption with the AuNP volume, which also intuitively scales with the AuNP concentration. Diluting the AuNP suspension shows a linear decrease in the absorption with constant absorption peak position (Figure~\ref{si:AuNPDilutionSeries} in the SI), verifying the assumption of a stable, dilute AuNP suspension (\Dh{} no plasmon coupling). Besides the primary peak, the absorption at low wavelength (around \SI{400}{nm}) decreases in case of precipitation. This is because the absorption in this range roughly scales with the gold content independently of AuNP shape, due to the interband transitions at wavelengths below \SI{380}{nm}. Interband transitions are excitations from valence to conduction band\cite{Blaber.2010}, and therefore, only dependent on the material properties, excluding NP-specific parameters~\cite{Merk.2014}, \Zb{} shape.

When the physisorbed citrate anions are replaced by 3-mercaptopropionic acid (MPA) -- as described in the synthesis section -- people usually refer to a `covalent' gold-sulfur bond that stabilizes the capping molecules at the gold surface~\cite{Hakkinen.2012}. Photos and absorption spectra of Figure~\ref{fig:AuNPOverview} demonstrate the successful capping exchange, and zeta potential measurements confirm a similar surface charge as the \mbox{L-Cit}. The \mbox{L-MPA} appear slightly larger than the L-Cit, yet,  the size statistics of the \mbox{L-MPA} is significantly lower, as they adsorb weaker to the TEM grids. This probably originates from weaker attraction of the MPA capping to the carbon film on the TEM grid. In addition, the exchange of the capping alters the direct electronic vicinity of the AuNPs, thereby also modifying the effective dielectric function of the surrounding medium, approximately accounting for a peak shift of \SI{2}{nm}\cite{Oliveira.2018}. In combination with the slightly higher averaged size, this leads to the small \mbox{red-shift} of the absorption maximum by \SI{4}{nm} in the present study. Besides the maximum position, the presence of the MPA capping broadens the peak due to the formation of a pseudo-insulating zone at the outer edge of the AuNP. Thus, excited plasmons experience additional damping\cite{Garcia.2005, Amendola.2017}. The peak height and low wavelength plateau are only slightly lower, indicating a minor AuNP loss during the capping exchange. 

Lastly, the absorption profile of the suspension changes when the AuNPs approach each other sufficiently close. In that case, the absorption profile develops a secondary absorption peak at higher wavelength compared to the primary absorption peak. Depending on the strength and the position of the secondary peak, this can occur as shoulder or broadening of the primary peak. For a relevant contribution of plasmon coupling to the absorption spectrum the inter-particle (surface-to-surface) distance should be smaller than \SI{20}{\percent} of the AuNP diameter~\cite{Amendola.2017, Lange.2012}. In the present work, this translates to roughly \SI{1}{nm} to \SI{2}{nm}. Therefore, this is mainly important when AuNPs are in close proximity such that the respective cappings interact, \Dh{} in case of aggregation. The absorption spectra taken over the range of \SI{72}{h} prove the stability of the pure AuNP suspension reference, which holds for months even.

The remaining discussion will separate the origins of AuNP aggregation into salt concentration-induced and specific-ion effects. Subsequently, the two sections deal with the influence of capping molecule and size, respectively.

\subsection{Effect of Salt Concentration on AuNP Aggregation}

The photo series' of the \mbox{L-Cit} and \mbox{S-Cit} (Figures~\ref{fig:photos_LCit}) reveal three different aggregation regimes ranging from stable suspensions via aggregation to fast destabilization with increasing salt concentration (\SI{10}{mM}--\SI{50}{mM}). In the case of \mbox{L-MPA} suspensions fully destabilize only at \SI{100}{mM} salt concentration. 

This transition originates from the screening effect of the salt ions and the size of the citrate capping. The Debye length $\Lambda_\mathrm{D}$ quantifies the range of electrostatic stabilization of a charged surface in a salted solution, and for monovalent electrolytes in water at room temperature calculates as~\cite{Israelachvili.2011}
\begin{equation}
	\Lambda_\mathrm{D} =\frac{\SI{0.304}{nm}}{\sqrt{ I[\si{M}] }} \, , 
\end{equation}
from the ionic strength of a solution $I$ in \si{M}. 
When calculating $\Lambda_\mathrm{D}$ for dilute AuNP suspensions (excluding screening by the AuNPs themselves) the results read: $\Lambda_\mathrm{D}= \SI{3.0}{nm}, \SI{1.9}{nm}, \SI{1.4}{nm}, \SI{1.0}{nm}$ for the respective salt concentrations (\SI{10}{mM}, \SI{25}{mM}, \SI{50}{mM}, \SI{100}{mM}). Thereby, the screening of any free citrate in suspension is neglected. Screening by either AuNPs or free citrate would further reduce the effective Debye length.
The size of a citrate anion in the capping layer can be estimated to \SI{0.8}{nm}~\cite{Grys.2020}. Consequently, the increase in salt concentration of \SI{10}{mM} to \SI{50}{mM} reduces the electrostatic stabilization range below twice the thickness of a citrate capping layer, enabling direct contact of two AuNPs. Furthermore, AuNPs in water attract each other at small separation due to van der Waals interactions \cite{Israelachvili.2011} and the physisorbed citrate capping is not able to prevent aggregation completely. as it can also facilitate bridging. This can happen either by hydrogen bonding of uncharged acid groups or by simultaneous adsorption at two approaching AuNP surfaces~\cite{Grys.2020}.

\subsection{Specific-Ion Effects on AuNP Aggregation}

\begin{figure}
	\includegraphics[width=0.85\textwidth]{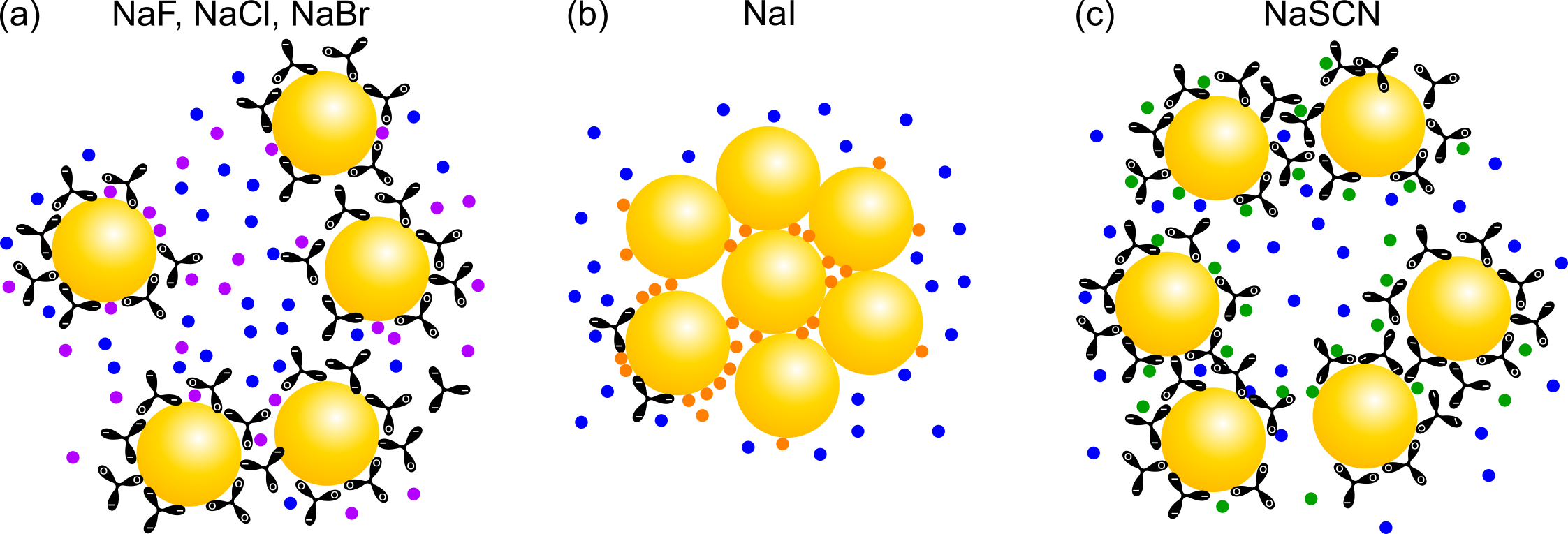}
	\caption{Schematic depiction of the different nanoscopic structure of AuNP aggregates induced by the various sodium salts. The large golden spheres with the surrounding black tripods represent the AuNPs and the citrate capping molecules, respectively. The sodium cations are blue and all anions in their plot color as before (\ch{Br-}: violet; \ch{I-}: orange; \ch{SCN-}: green). (a) Non-distinct aggregation of AuNPs with strongly varying inter-particle distance, induced by \ch{F-}, \ch{Cl-} or \ch{Br-} (\ch{Br-} shown as example). (b) Removal of citrate capping from AuNPs by \ch{I-} and subsequent formation into solid gold structures. (c) AuNP pair formation with narrow distribution of inter-particle distance by adsorption of \ch{SCN-} to the capping layer and subsequent bridging.}
	\label{fig:discussionScheme}
\end{figure}

The photo series' of all AuNP types reveal ion-specific effects, which can be discriminated by the color change of the suspension and in greater detail by the absorption spectra. As mentioned in the introduction, ion-specific behavior is a well-known concept and the strength of ion-specific effects usually orders along the Hofmeister series, \Dh{} $\ch{F^-} < \ch{Cl^-} < \ch{Br^-} < \ch{I^-} < \ch{SCN^-}$, with respect to their tendency to adsorb to interfaces. Figure~\ref{fig:discussionScheme} presents a scheme of the ion-specific differences discussed in the following paragraphs. According to the results we distinguish three categories of ions: 1) \ch{F-}, \ch{Cl-}, \ch{Br-}; 2) \ch{I-}; and 3) \ch{SCN-}.

\noindent\textbf{1) \ch{F-}, \ch{Cl-}, \ch{Br-}:} At \SI{10}{mM} sodium salt concentration the primary absorption peak is almost unaffected over time, while at \SI{25}{mM} the absorption spectra reveal a decreasing reduction of the primary absorption peak over time from NaF to NaBr (Figure~\ref{fig:timeMatrix_LCit}). For the latter, the reduction of the peak is almost negligible. The more chaotropic \ch{Br^-} ions are expected to adsorb at the AuNP surface stronger than the other two anions leading to an increase of the AuNPs' negative surface charge which counteracts the electrostatic screening by the salt. For higher salt concentration (\SI{50}{mM}) the general trend continues, showing slower precipitation for NaBr compared to NaCl and NaF. However, the high salt concentration and resulting electrostatic screening lead to simultaneous formation of large aggregates with a wide distribution of inter-particle distance between the individual AuNPs (Figure~\ref{fig:discussionScheme} (a)). Thereby, the AuNPs can form a large porous structure (TEM images, Figure~\ref{fig:TEMseries_LCit}) that generates the approximately constant absorption profiles, which was observed for similar systems at higher salt concentrations and labeled `gold sponge'~\cite{Zhang.2014}. 

\noindent\textbf{2) \ch{I-}:} In contrast, the photo series' and absorption spectra of suspensions containing NaI already clearly demonstrate a very different behavior compared to the above discussed sodium halides. The Hofmeister series would indicate even stronger interface affinity of the \ch{I^-} ions, and thereby, higher colloidal stability. Indeed, it is known that the affinity of \ch{I^-} to gold surface is very high, even sufficiently strong to stabilize uncapped AuNPs in suspension, where cosmotropic \ch{F^-} leads to AuNP aggregation~\cite{Merk.2014}. Adsorption of \ch{I-} ions to the gold surface changes the electronic environment of the AuNPs resulting in a red-shift of the primary absorption peak. However, both the \ch{I-} and the citrate anions prefer to adsorb at flat gold interfaces~\cite{Magnussen.2002,Perfilieva.2019}, which leads to competitive detachment of the physisorbed citrate capping. Subsequently, the AuNPs are not protected against close contact interaction anymore. The strong van der Waals-attraction of AuNPs~\cite{Israelachvili.2011} then leads to successive aggregation of the partly capped AuNPs (Figure~\ref{fig:discussionScheme} (b)). This is in contrast to the trend of NaF to NaBr, where the increasing ion adsorption leads to increased AuNP stabilization due to increased negative surface charge. Nevertheless, the adsorption of \ch{I^-} still initially increases the negative surface charge as well, which accounts for the small red-shift of the absorption peak immediately after mixing. In parallel, the competitive desorption of citrate anions takes time, which is even further amplified by the fact that AuNPs need to come close to attract each other \latin{via} van der Waals interaction but the suspension is very dilute ($\approx \SI{5E-6}{v\per v}$). Therefore, immediately after mixing NaI induces a similarly weak response as NaBr. 

The aggregation of partly capped AuNPs leads to direct contact of the AuNPs during aggregation, thus, increasing the effective size of the AuNPs. As discussed above, increasing the AuNP size correlates with a \mbox{red-shift} of the absorption maximum~\cite{Kimling.2006}. Therefore, the continuous \mbox{red-shift} of the primary peak in NaI solutions during the first \SI{12}{h} can be attributed to the successive increase of the effective AuNP size. As AuNPs grow further, the frequency of aggregation events decreases with the decreasing AuNP number density, while the precipitation of AuNP aggregates accelerates. Consequently, precipitation will inhibit observation of further AuNP growth after roughly \SI{12}{h} at a primary peak maximum around \SI{540}{nm}. However, the direct correlation between absorption maximum and increasing effective AuNP size requires conductive contact~\cite{Amendola.2017}; otherwise the spectrum would feature a secondary absorption peak due to plasmon coupling (SCN-paragraph below). The whole aggregation process of (partly) uncapped AuNPs by NaI involves a subsequent step, as \ch{I^-} can cause fusion of AuNPs in suspension~\cite{Cheng.2003}. This process enables the transition from porous AuNP aggregates to large solid structure many times the size of the initial AuNPs (roughly \SI{10}{nm}). Comparing the observed absorption peak position of \mbox{L-Cit} with absorption spectra of larger AuNPs~\cite{Kimling.2006, NanopartzInc..09.02.2024} yields a diameter around \SI{75}{nm} to \SI{80}{nm}. The TEM images (Figures~\ref{fig:TEMseries_LCit},~\ref{fig:TEMseries_SCit},~\ref{fig:TEMseries_MPA}) indicate even larger aggregate sizes around \SI{100}{nm} or more. The differences likely stem from the fact that the fused aggregates are less spherical and more polydisperse (cf. TEM images) than the freshly synthesized larger AuNPs (as referenced). Furthermore, \ch{I-} should be incorporated in the aggregate as the fusion requires \ch{I-} adsorption, which alters the absorption profile by changing the electronic environment and internal structure of the AuNPs. Besides, AuNPs may stack on TEM grids which prevents clear distinction of two AuNPs in some cases. Taking into account all these effects, the values agree quite well.
    
Some fusion of AuNPs into larger structures is also observable for NaCl, but the resulting structures remain around \SI{50}{nm}. Similar signs of fusion have been observed for sodium halides before, but a much higher salt concentrations around \SI{200}{mM}~\cite{Zhang.2014} (compared to \SI{50}{mM} here). There, all sodium halides induce that effect but it was much stronger for NaCl than NaF and NaBr. Obviously, a replacement of citrate comparable to the \ch{I-}-case does not take place in presence of NaCl, since the primary absorption peak is neither shifted nor does it remain a well-defined peak. In contrast, the spectra and images indicate aggregation of AuNPs in the `sponge-like' structure, leading to the broad absorption plateau that we discussed above. We speculate that the reason for NaCl-induced AuNP fusion lies in the nature of the AuNP synthesis, since \ch{Na^+} and \ch{Cl^-} ions are already present during the AuNP synthesis as counter-ions of the used gold salt and sodium citrate. The surface of synthesized AuNPs, esp. the presented crystal plane depends strongly on the ion type of present salts~\cite{Ha.2007, Morandi.2020}. Therefore, the outer surface may promote the adsorption of some NaCl enabling the fusion of AuNP at certain sites on the AuNP surface that are not protected by citrate capping. This would explain why the fused aggregates for NaCl are more elongated and smaller than their NaI counterparts. However, the underlying mechanism is still unclear and requires further investigation. Nevertheless, the appearance of the wide absorption plateau is a strong argument for a high fraction of remaining citrate molecules to separate the AuNPs from each other and retaining a porous structure. 

\noindent\textbf{3) \ch{SCN-}:} The most chaotropic - and therefore most surface affine - anion in the present series is \ch{SCN^-}. As for NaI, the aging absorption spectra of NaSCN-containing suspensions reveal distinct changes in the absorption profiles. In contrast to NaI, the position of the primary peak is not specially affected and decreases similar to the halides, \Zb{} NaF. Therefore, we conclude that the capping is not replaced by \ch{SCN-}. However, during aging a distinct secondary absorption peak forms roughly at a wavelength of \SI{650}{nm}. This is due to the plasmon coupling of AuNPs in close vicinity to each other, as discussed earlier. This is the same mechanism that leads to absorption plateau at higher wavelength for \SI{50}{mM} NaF, NaCl, and NaBr. However, for NaSCN a distinct peak appears (instead of a broad shoulder) indicating a relatively narrow distribution of the inter-particle distance, which is proven by the TEM images. Thus, this secondary peak suggests that small aggregates of two or a few AuNPs form (Figure~\ref{fig:discussionScheme} (c)), although the successive \mbox{red-shift} of the secondary peak indicates a slight growth of the aggregates over time. The same is visible in the TEM images (Figure~\ref{fig:TEMseries_LCit}) that indicate the formation of mostly mono- or bilayers of AuNPs on the TEM grid for \ch{SCN-}. Since the formation of two-dimensional aggregates in suspension is highly unlikely, we conclude that only pairs and small aggregates are present in suspension, that assemble into the large two-dimensional structure during adsorption to the TEM grid. In contrast, all the halides form clear three-dimensional aggregates that consist of multiple layers of AuNPs in each direction. This is the structure one would expect for aggregation of nanoparticles in suspension. The reason for the formation of the small aggregates in the presence of \ch{SCN-} lies in their strong surface affinity. This allows them to adsorb at the outer surface of the capping layer of the AuNPs, besides the adsorption at the bare gold surface. Therefore, they don't replace the citrate capping at the surface, and further increase the negative surface charge of the AuNPs (similar to \ch{Br-}). However, the AuNPs form pairs or small aggregates to optimize the absence of water near the chaotropic \ch{SCN-} anions, effectively leading to AuNP bridging. However, as the \ch{SCN-} are very bulky they prohibit direct AuNP contact. Thus, the absorption spectra feature relatively stable suspensions and a secondary absorption peak due to plasmon-coupling of neighboring AuNPs, while the TEM images show well-separated AuNP assemblies.

Explanation of the non-systematic effects from NaF to NaSCN along the Hofmeister series involves consideration of two adsorption interfaces that are present for the various ions, as schematically depicted in Figure~\ref{fig:discussionScheme}. Anions can either adsorb to the capping layer covering a AuNPs or the bare AuNP surface, competing for space with the capping molecules. Therefore, we conclude that the difference between NaI and NaSCN mainly results from the specific affinity of \ch{I^-} to the gold surface compared to the citrate capping, while NaSCN can adsorb to the citrate capping, leaving the capping layer sufficiently intact, but instead bridging two AuNPs due to the very chaotropic nature.

\subsection{Effect of AuNP Capping}

The photo series of \mbox{L-MPA} clearly shows the enhanced stability of the suspension compared to \mbox{L-Cit} as a direct consequence of the stronger bound capping. Only at a salt concentration of \SI{100}{mM} all suspensions experience fast and complete destabilization by AuNP aggregation. Interestingly, there is only a small ion-specific variation of the absorption spectra during aggregation except for NaI. For all the other salts, the spectra develop a secondary absorption peak that manifests in a high wavelength shoulder of the primary absorption peak (similar to \mbox{L-Cit} with NaSCN). The formation of AuNP pairs or small aggregates results from screening of the electrostatic stabilization by the high salt concentration. Since the AuNP separation is stronger for MPA compared to citrate, the AuNPs initially form smaller aggregates, consisting of well separated AuNPs as visible in the TEM images (Figure~\ref{fig:TEMseries_MPA}). Only subsequently, these aggregates grow in size and appear less distinct leading to an approximately constant absorption profile.

The ion-specific effects of NaI appear very similar to \mbox{L-Cit}: at \SI{25}{mM} the absorption red-shifts  gradually over \SI{72}{h} (up to around \SI{550}{nm}) with simultaneous peak broadening; at \SI{50}{mM} and \SI{100}{mM} the suspension destabilizes very fast, even avoiding the full red-shift. The strength and process of the AuNP aggregation indicates that the \ch{I-} ions are able to remove the capping similar to the citrate ions, despite the bond usually being referred to as `covalent' due to its strength. Recent simulations for example estimate the individual bond strength of thiols around \SI{1.1}{eV}~\cite{Inkpen.2019}. However, measurements of similar thiol-based capping agents that cannot form covalent bonds show identical bond properties~\cite{Inkpen.2019,Pacchioni.2019}. Although the simulated values need to be treated with care (as stated by the authors), the results raise doubts about the covalent character of the Au-thiol bond. Measurements of the association of \ch{I-} ions to Au surfaces reveal adsorption energies of the same order of magnitude, for instance around \SI{0.7}{eV} to a \{111\} Au surface at \SI{1}{mM} NaI concentration~\cite{Deakin.1988}. This is already reasonably close to the binding energy of the thiol-capping agents. Furthermore, the adsorption energy of \ch{I-} ions to a gold surface increases with the salt concentration in solution, roughly by \SI{100}{\milli eV} per decade of NaI concentration~\cite{Ueno.1999}. In contrast, the bond strength of the thiol-capping to the gold surface decreases with shorter carbon chains. Therefore, the combination of the short chain thiol MPA and a reasonably high concentration of NaI in the AuNP suspensions leads to AuNP aggregation. 

The TEM images of AuNP suspensions in Figure \ref{fig:TEMseries_MPA} support these findings. Except NaI, TEM images for all salt show predominantly monolayer assembly of AuNPs indicating AuNP assembly only during adsorption to the TEM grid as opposed to aggregation in suspension (as discussed above for NaSCN and L-Cit). The difference to the \mbox{L-Cit} samples originates from the increased stability of the \mbox{L-MPA} suspension that already showed in the absorption spectra. In the presence of NaI AuNPs experience fusion again. However, the extent of the fusion is strongly retarded compared to the \mbox{L-Cit} (cf. Figure \ref{fig:TEMseries_LCit}). This delay can be attributed to the much stronger bond of MPA to the gold surface obstructing the capping removal.

\subsection{Effect of AuNP Size}

Comparing to \mbox{L-Cit}, \mbox{S-Cit} suspensions appear more stable (Figure~\ref{fig:photos_SCit}), which is due to lower van-der-Waals attraction for the \mbox{S-Cit} AuNPs: the van der Waals attraction energy for two AuNPs reads\cite{Israelachvili.2011}
\begin{equation}
	E_\mathrm{vdW}=-\frac{A}{12D} R \ , 
\end{equation}
with the Hamaker constant $A$ (for two gold surfaces in water), the inter-particle distance $D$, and the AuNP radius $R$. Therefore, the attraction between smaller AuNPs is reduced compared to larger ones. However, the mixtures simultaneously exhibit more distinct ion-specific effects over all three salt concentrations, with three outstanding effects.

First, NaBr stabilizes the AuNP suspension even at \SI{50}{mM}. This matches the trend for \mbox{L-Cit} of NaBr inducing the slowest and weakest response. This effect occurs more pronounced here, which results from the higher availability of adsorption surface (smaller AuNPs at identical gold salt concentration, cf. `Materials'). In addition to the geometrical increase of AuNP surface, the density of citrate molecules at the gold surface is assumed to be lower as indicated by the lower Zeta-potential, which leads to even more effective gold surface compared to \mbox{L-Cit} suspensions. More adsorption sites for \ch{Br-} ions lead to an increase in total surface charges of the S-Cit, while simultaneously decreasing free salt ions in solution that can participate in screening. Both of these effects lead to increased stability of the suspension. The strong adsorption of \ch{Br-} ions leads to a red-shift of the absorption peak (Figure~\ref{si:timeMatrix_SCit}) due to a change of the effective refractive index in direct vicinity of the AuNPs, as also observed by other authors~\cite{Zhang.2014}. As discussed above, \mbox{S-Cit} suspensions containing \SI{100}{mM} salt show fast precipitation for all salts (incl. NaBr) similar to \mbox{L-Cit} at \SI{50}{mM} salt concentration. The available surface scales inversely with the AuNP diameter, meaning the \mbox{S-Cit} suspensions provide roughly double the total particle surface compared to the \mbox{L-Cit} suspensions. That suggests that there is no change in the fundamental effects but only a shift of the destabilization by NaBr to a higher concentration, where saturation of the increased surface is achieved and repulsion is fully screened. This is also supported by the fact that the aggregation and precipitation dynamics are delayed for all anions except \ch{I-} (Figure~\ref{si:timeMatrix_SCit}).

Secondly, NaF produces a change to a brownish coloring at \SI{25}{mM} and \SI{50}{mM} over time for \mbox{S-Cit} originating from a broad high-wavelength shoulder, while NaF does not change the primary peak position or shape for \mbox{L-Cit}. This results from a stronger screening where little to no adsorption of \ch{F-} ions to the AuNPs is expected. NaF produced a stronger salting-out for \mbox{L-Cit} suspensions as well (compared to NaCl or NaBr), and the strong increase in AuNP number density in suspension (roughly by a factor of 10) enhances the effect, analogue to the increased stability for NaBr.

The third and most prominent anomaly occurs for NaI. NaI starts to replace the citrate capping already at the lowest salt concentration (\SI{10}{mM}) within \SI{24}{h}, while this was barely the case for the \mbox{L-Cit} even after \SI{72}{h}. Here the absorption not only red-shifts as observed before, but also the peak intensity increases while the \mbox{low-wavelength} plateau decreases significantly (Figure~\ref{fig:timeMatrix_SCit}). Thereby the peak becomes more pronounced, since the relative peak width decreases. At \SI{25}{mM} the process is faster and leads to complete precipitation over \SI{72}{h}. The narrowing of the absorption peak is an intrinsic effect of the AuNP fusion by NaI opposed to aggregation of AuNPs that keep an internal structure. As discussed at the start of the chapter, the width of the primary absorption peak is dominated by the polydispersity of the AuNP diameter and only slightly depends on the mean AuNP size. Therefore, the formation of AuNP aggregates of various sizes generally implies a broadening of the absorption peak, which is observable at high NaI concentrations (as it was for \mbox{L-Cit}). In addition to the fusion of two AuNPs, it has been observed before that \ch{I-} may facilitate a leeching of gold from one AuNP to another, similar to `Ostwald-ripening'~\cite{Cheng.2003}. Therefore, the fused AuNP aggregates can transform into solid particles of much larger size, as visible in TEM images, when the destabilization process takes many hours or days (as for \SI{10}{mM} and \SI{25}{mM} NaI). As discussed for \mbox{L-Cit} suspension, TEM images and the primary absorption position shift indicate aggregate diameters well above \SI{50}{nm}, building one aggregate out many hundreds to a few thousand AuNPs of initial diameter around \SI{5}{nm}. Thus, one aggregate likely consists of a good representation of the size distribution of the \mbox{S-Cit} suspension. Consequently, the resulting size dispersity of the fused AuNP aggregates can become less broad than the distribution of initial AuNPs, which leads to a narrowing of the resulting peak. This process is strongly limited by the distance between AuNPs, \Dh{} the AuNP number density. With progressing aggregation and fusion, fewer AuNPs are available in each others reach, resulting in a final position of the primary absorption peak around \SI{540}{nm}. Since the amount of gold is identical, the final peak positions of \mbox{L-Cit} and \mbox{S-Cit} suspensions agree quite well. During the whole process the low-wavelength plateau decreases monotonically showing that precipitation of AuNP aggregates (probably larger ones) superimposes the above described process. For all NaI concentrations precipitation is also clearly visible for the primary peak once the red-shift completes.

In addition to the stronger iodide-specific response, the effects related to iodide-gold interaction happen faster compared to the \mbox{L-Cit}. This probably results from a lower density of citrate anions at the AuNP surface, as discussed for NaBr already. The affinity of the ions to the gold surface depends on the smoothness of the surfaces, which results from the respective crystal plane (\{111\} smoother than \{110\} smoother than \{100\}). Both anions, citrate~\cite{Perfilieva.2019} and \ch{I-}~\cite{Magnussen.2002}, prefer smooth AuNP surfaces. Thus, the citrate anions will predominantly sit at \{111\}-facets, and therefore, be easier removed by \ch{I-}.

Absorption spectra of suspensions containing NaSCN at \SI{25}{mM} show barely any response except for the small \mbox{red-shift} of the primary absorption peak that occurs for all salts, as discussed above. At higher salt concentration (\SI{50}{mM}) the suspension is destabilized -- again slower compared to \mbox{L-Cit}. Since \ch{SCN-} ions adsorb to the capping layer as well as gold surface, there are no specific differences visible compared with the \mbox{L-Cit} suspensions, aside the slower and weaker destabilization due to the increase in overall AuNP surface and reduced van der Waals attraction. The spectra reveal the beginnings of a secondary absorption peak after \SI{6}{h} and \SI{12}{h} (partly out of measure range already). The peak is so broad and poorly defined that it appears as a shoulder of the primary peak instead of a separate one. This results from the less distinct initial absorption peak as a consequence of the higher size distribution of the \mbox{S-Cit}.

\section{Conclusion}
\label{conclusion}
The study focuses on the concentration-dependent and ion-specific aging (up to \SI{72}{h}) of three types of AuNPs, starting from \SI{11}{nm}, citrate-capped AuNPs, and subsequently varying the size (\SI{5}{nm}) and the capping type (3-mercaptopropionic acid), respectively. The employed sodium salts vary the anion along the Hofmeister series (NaF, NaCl, NaBr, NaI, NaSCN). The ion-specific AuNP aggregation and suspension stability are illustrated by TEM, absorption spectroscopy, and photography. 
The salt concentration covers the full range from charge-stabilized suspensions (\SI{10}{mM}) to fast destabilization by electrostatic screening (\SI{50}{mM}). At an intermediate salt concentration of \SI{25}{mM} the AuNPs already exhibited some aggregation but the AuNP suspensions mainly retained their stability. Decreasing the AuNP diameter stabilizes the suspensions against AuNP aggregation due to the lowered van der Waals attraction, and exchanging the citrate capping to covalently bound MPA strongly suppresses AuNP aggregation, as well. 
At a salt concentration of \SI{25}{mM} ion-specific aging becomes visible (around \SI{12}{h}) in the form of color changes of the originally red AuNP suspensions. Along the Hofmeister series the suspensions stability increases from NaF to NaBr (`salting-in') due to the increasing adsorption of anions to the negatively charged AuNPs. NaI exhibits even higher surface affinity to the AuNP surface leading to (competitive) desorption of the citrate capping. Subsequently, AuNPs aggregate into large clusters leading to a break-down of suspension stability. This change in stabilizing behavior from NaBr to NaI is surprising with respect to the Hofmeister series, which can be explained by specific binding of \ch{I-} to the gold surface resulting in citrate capping removal. Continuing to the more chaotropic \ch{SCN-} ion, the suspensions experience weaker destabilization compared to NaI, underlining the non-monotonous behavior. In contrast to \ch{I-}, the \ch{SCN-} ions adsorb to the citrate capping molecule, thereby not inducing the capping removal, but instead leading to bridging between AuNPs due to the strong chaotropic character of \ch{SCN-}. Therefore, NaSCN leads to AuNP aggregates that retain an inner substructure, while NaI causes large, solid structures by fusion of the individual AuNPs.
Interestingly, \mbox{S-Cit} suspensions exhibit a broad range of ion-specific effects, despite the weaker response to the salt concentration. The reason here are the higher particle number density and probably easier access of the ions to the AuNP surface. The `salting-out' to `salting-in' transition occurs more pronounced, manifesting in decreased stabilization for NaF and improved stabilization by NaBr compared to the \mbox{L-Cit} behavior. Most importantly, the destabilizing effect of NaI occurs very distinct, already clearly visible at \SI{10}{mM} salt concentration. 
This effect can even be observed for \mbox{L-MPA} suspensions, which otherwise proved relatively robust against AuNP aggregation and corresponding suspension destabilization. Nevertheless, the \ch{I-} ions are able to replace the `covalently' bound MPA capping, even though the process is slower due to the increased capping bond. 
When using AuNPs in saline environments, esp. the effect of \ch{I^-} ions needs to be taken into account to ensure appropriate functions for composite materials. Otherwise, \ch{I^-} can attack stabilizing agents possibly leading to additional (unwanted) effects in complex multi-component systems~\cite{Wang.2020}. The presently discussed MPA capping only has a very short hydrocarbon chain (two methylene units) between the anchor point and the carboxy-group. Further investigations of longer chain thiol agents should evaluate if a better shielding of the gold surface from \ch{I-} ions can prevent the removal of the MPA molecules.
Aside from pure suspensions, the knowledge gained for the colloidal stability of AuNP suspension might help to tailor composite materials made of AuNPs embedded in a polymer matrix.

\begin{acknowledgement}

\end{acknowledgement}
Funding from the German Science Foundation is acknowledged: R.v.K. is member of the RTG 2516 (Grant No. 405552959) and P.R. is the recipient of a doctoral position within the RTG 2516 program.
\subsection{Conflict of Interest}
The authors have no conflict to declare.

\subsection{Author Contributions}
Conceptualization, P.R., R.v.K.; formal analysis, P.R.; investigation, P.R., A.S.;  writing--original draft preparation, P.R.; writing--review and editing, P.R., R.v.K.; visualization, P.R.; supervision, R.v.K.; project administration, R.v.K.; funding acquisition, R.v.K. All authors have read and agreed to the published version of the manuscript.

\newpage
\bibliography{literature_finalDraft.bib}

\newpage
\appendix
\begin{suppinfo}
\section{Supporting Information (SI)}
	\setcounter{figure}{0} 
	\renewcommand{\thefigure}{S\arabic{figure}}
	\pagenumbering{Roman}
	\begin{figure}
	\includegraphics[width=0.74\textwidth]{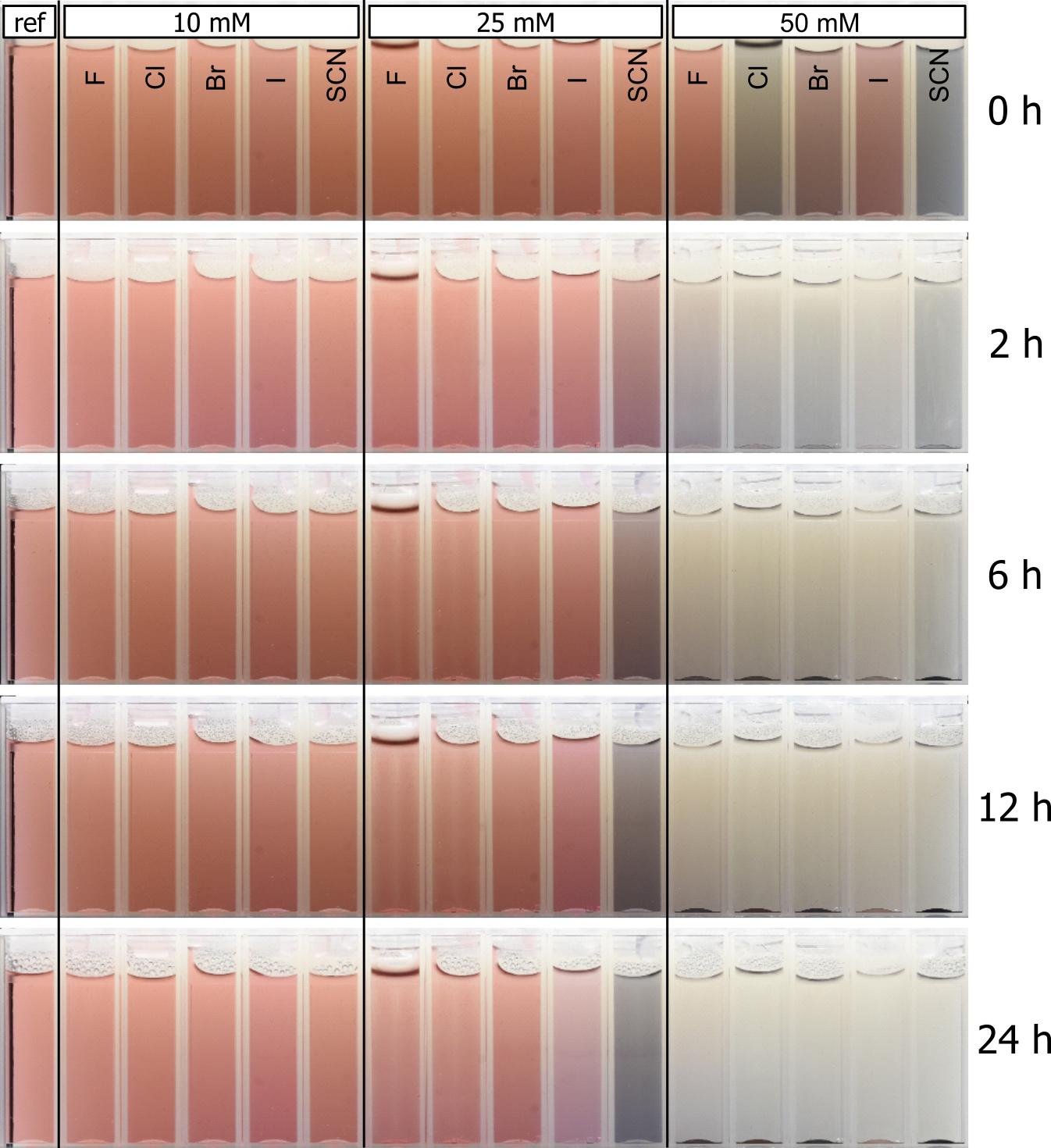}
	\caption{Camera snapshots of \mbox{L-Cit} suspensions mixed with sodium salts, with anions varying along the Hofmeister series: NaF, NaCl, NaBr, NaI, NaSCN, separated by rows. The salt concentration varies from \SI{10}{mM} to \SI{100}{mM}.}
	\label{si:photos_LCit}
\end{figure}

\begin{figure}
	\includegraphics[width=0.75\textwidth]{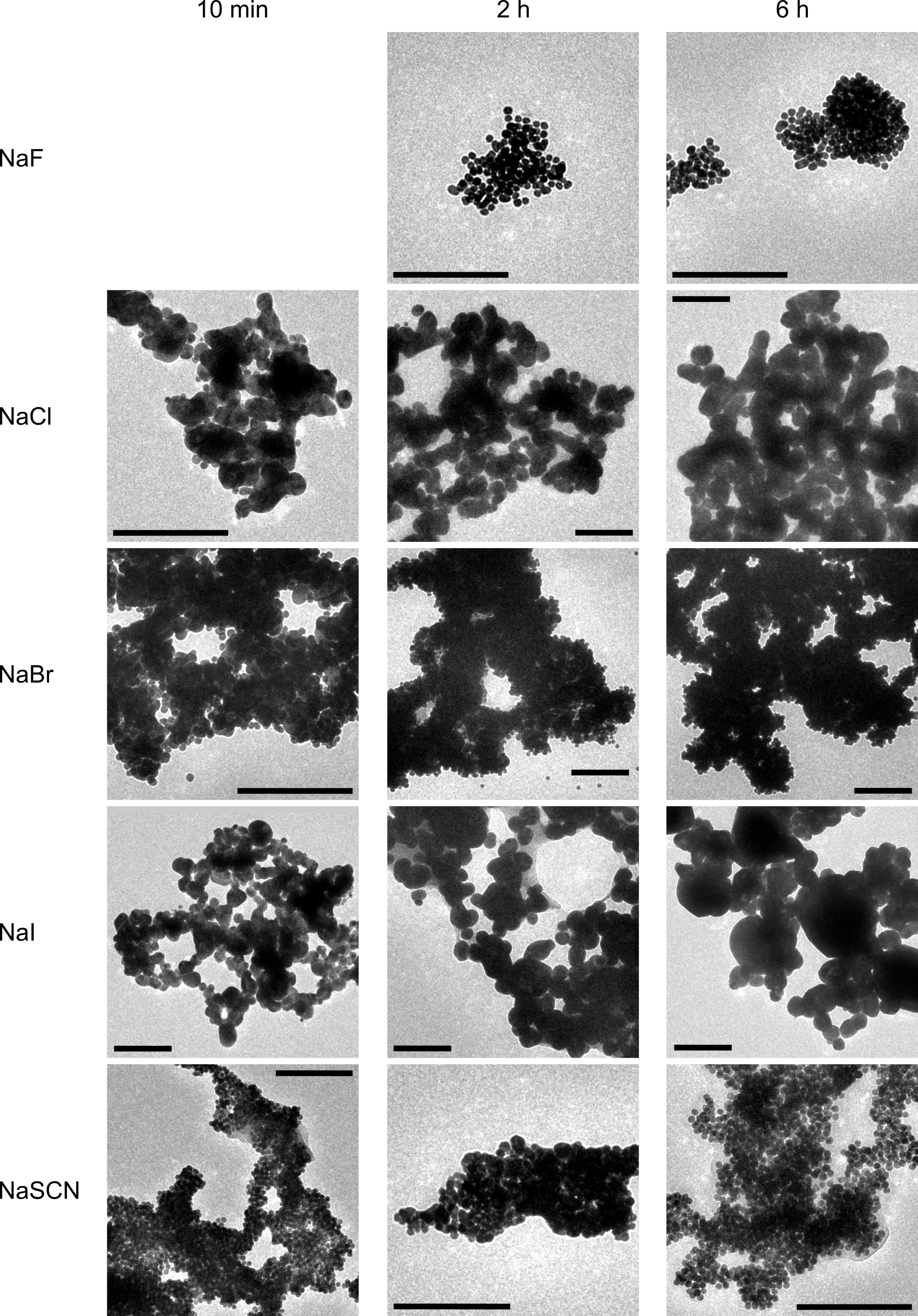}
	\caption{Aging time series for \mbox{L-Cit} suspensions \SI{10}{min}, \SI{2}{h}, \SI{6}{h} after addition of \SI{50}{mM} of the respective salt. Different anions ordered in rows along the Hofmeister series as before: F, Cl, Br, I, SCN. The scale bars in all images represent \SI{200}{nm}. Aggregation took place in suspension and aggregates assembled on a TEM grid for only \SI{5}{min}.}
	\label{si:TEMseries_LCit}
\end{figure}

\begin{figure}
	\includegraphics[width=0.95\textwidth]{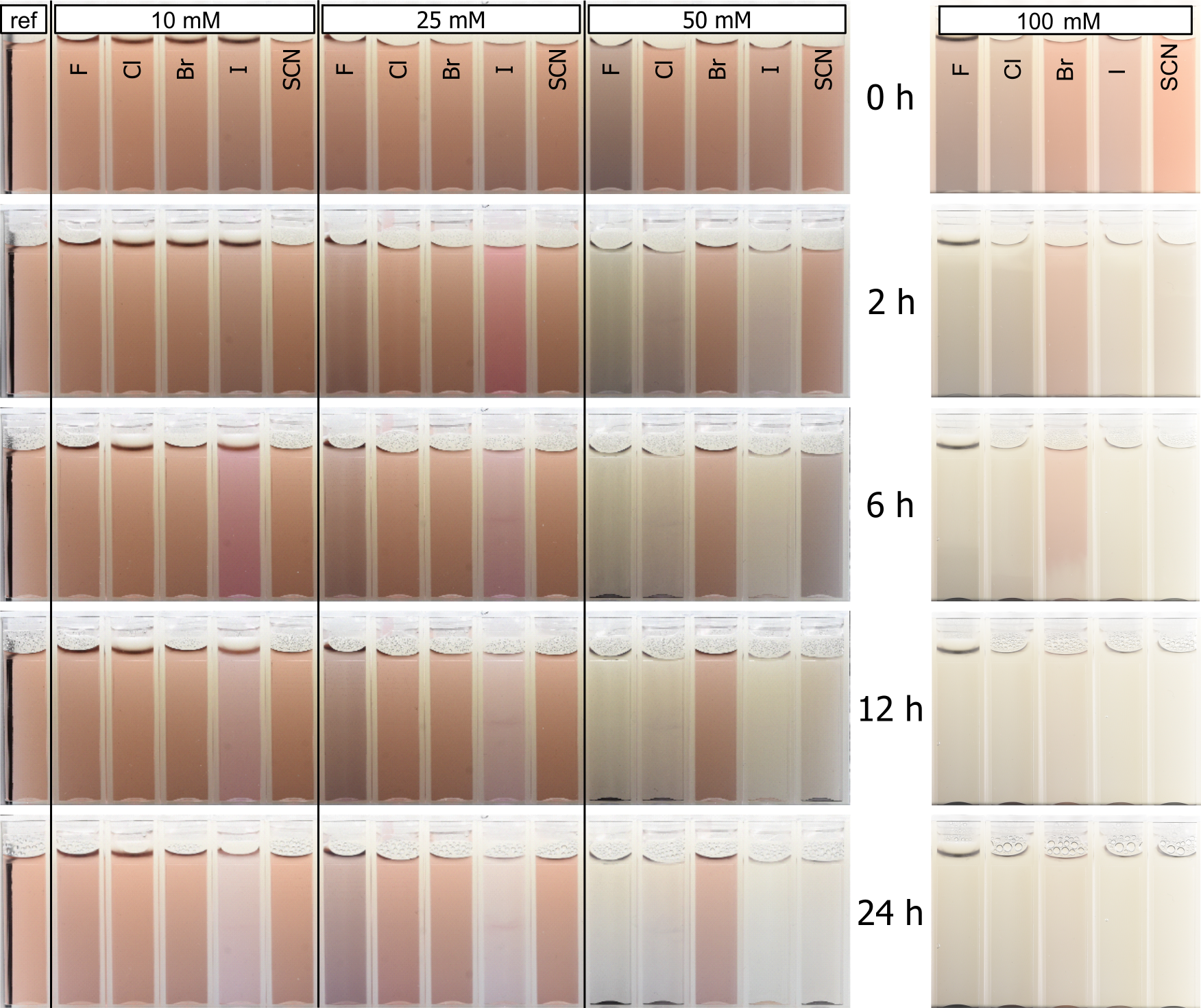}
	\caption{Camera snapshots of \mbox{S-Cit} suspensions mixed with sodium salts, with anions varying along the Hofmeister series: NaF, NaCl, NaBr, NaI, NaSCN, separated by rows. The salt concentration varies from \SI{10}{mM} to \SI{100}{mM}.}
	\label{si:photos_SCit}
\end{figure}

\begin{figure}
	\includegraphics[width=0.95\textwidth]{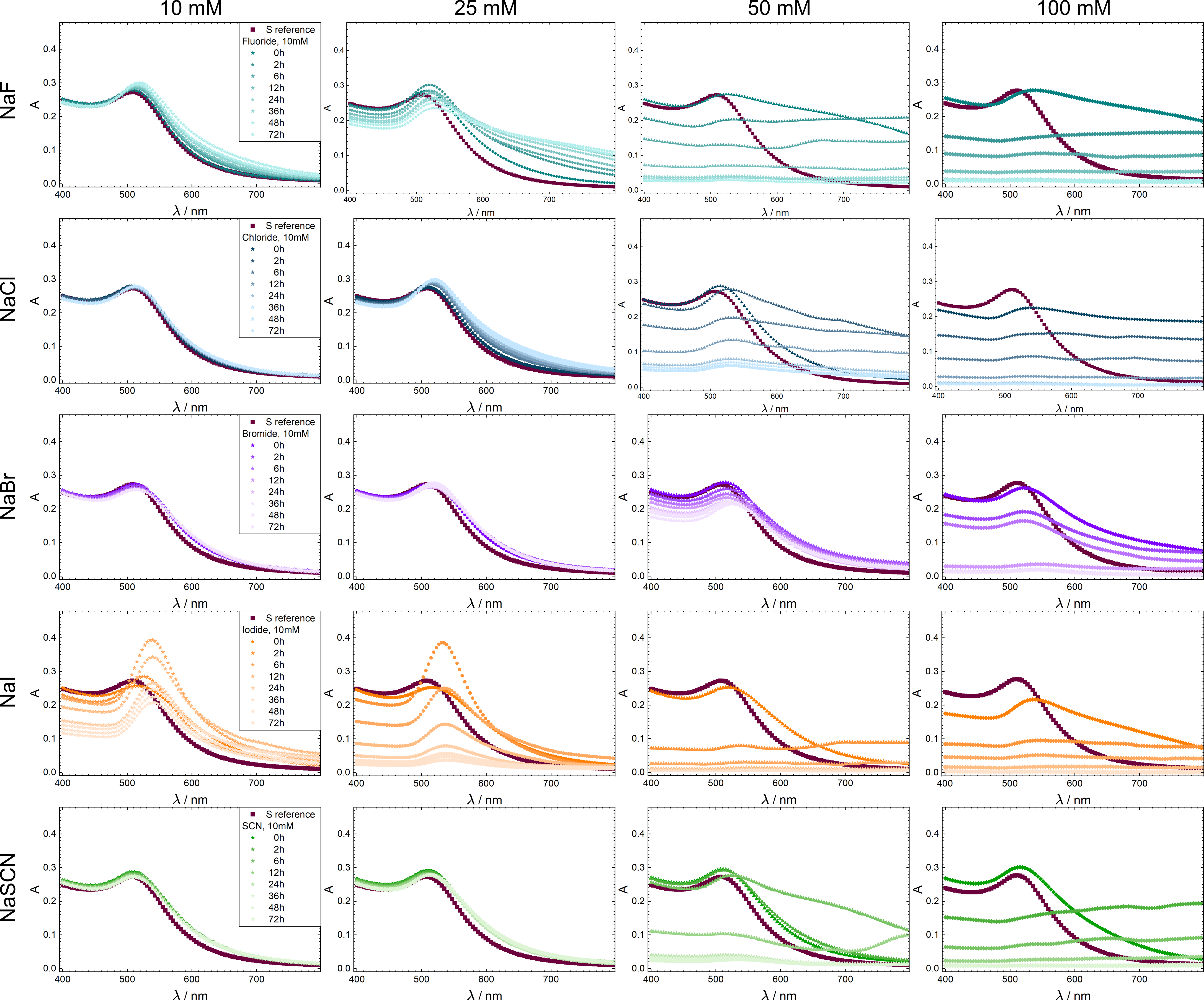}
	\caption{Absorption spectra of \mbox{S-Cit} suspensions mixed with sodium salts, with anions varying along the Hofmeister series: NaF, NaCl, NaBr, NaI, NaSCN, separated by rows. The salt concentration varies from \SI{10}{mM} to \SI{100}{mM}.}
	\label{si:timeMatrix_SCit}
\end{figure}
%

\begin{figure}
	\includegraphics[width=0.95\textwidth]{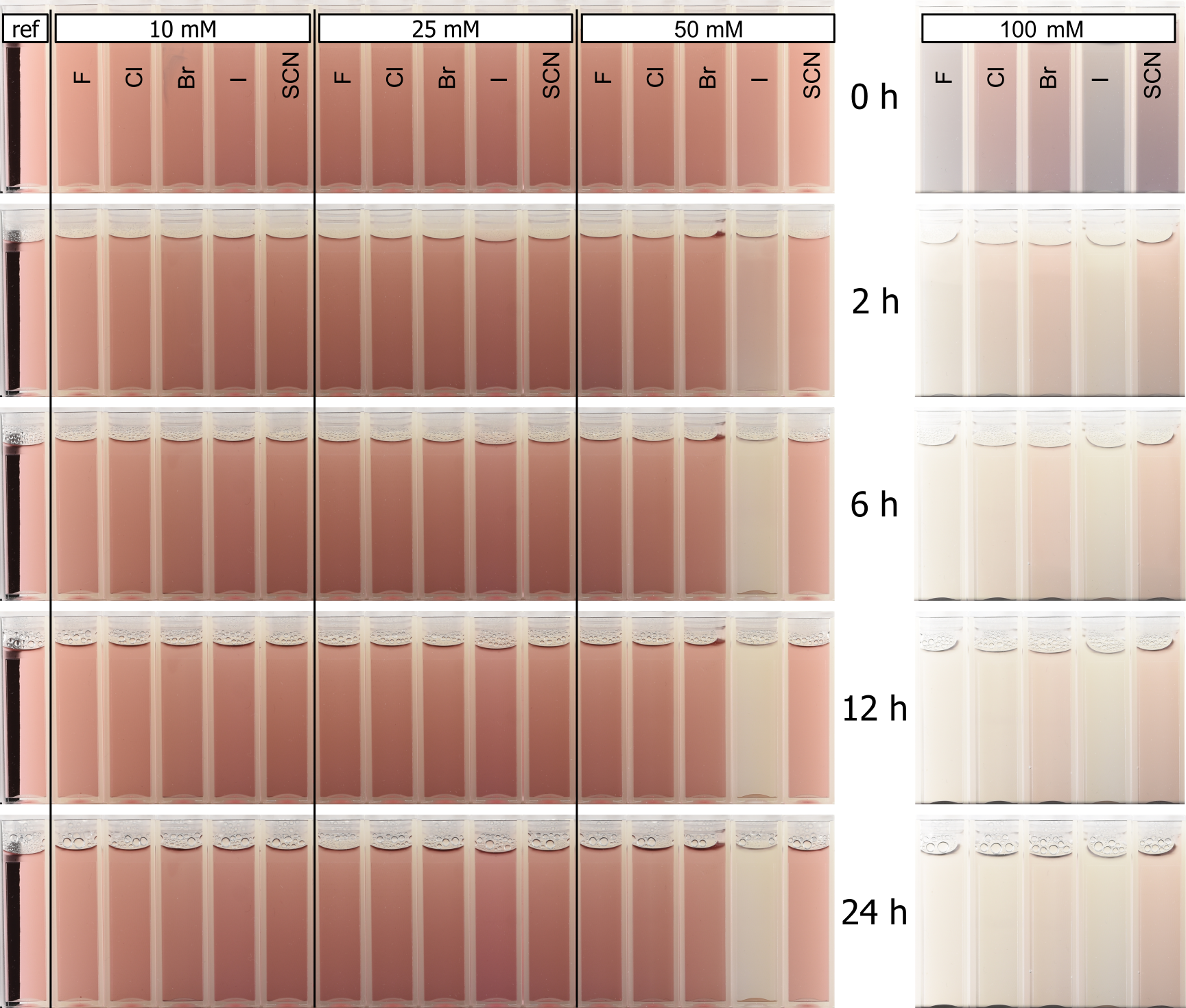}
	\caption{Camera snapshots of \mbox{L-MPA} suspensions mixed with sodium salts, with anions varying along the Hofmeister series: NaF, NaCl, NaBr, NaI, NaSCN, separated by rows. The salt concentration varies from \SI{10}{mM} to \SI{100}{mM}.}
	\label{si:photos_LMPA}
\end{figure}

\begin{figure}
	\includegraphics[width=0.95\textwidth]{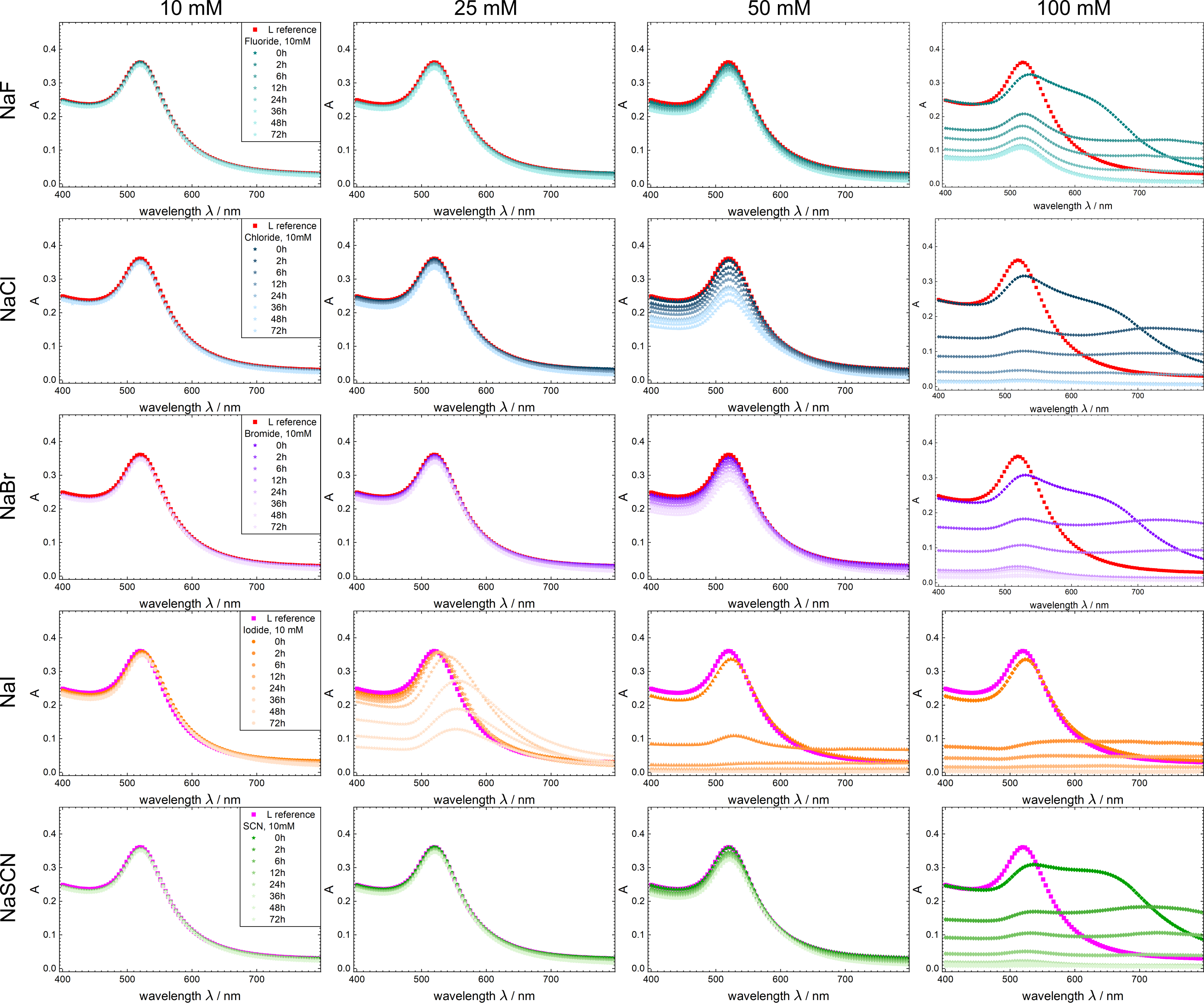}
	\caption{Absorption spectra of \mbox{L-MPA} suspensions mixed with sodium salts, with anions varying along the Hofmeister series: NaF, NaCl, NaBr, NaI, NaSCN, separated by rows. The salt concentration varies from \SI{10}{mM} to \SI{100}{mM}.}
	\label{si:timeMatrix_LMPA}
\end{figure}

\begin{figure}
	\includegraphics[width=0.75\textwidth]{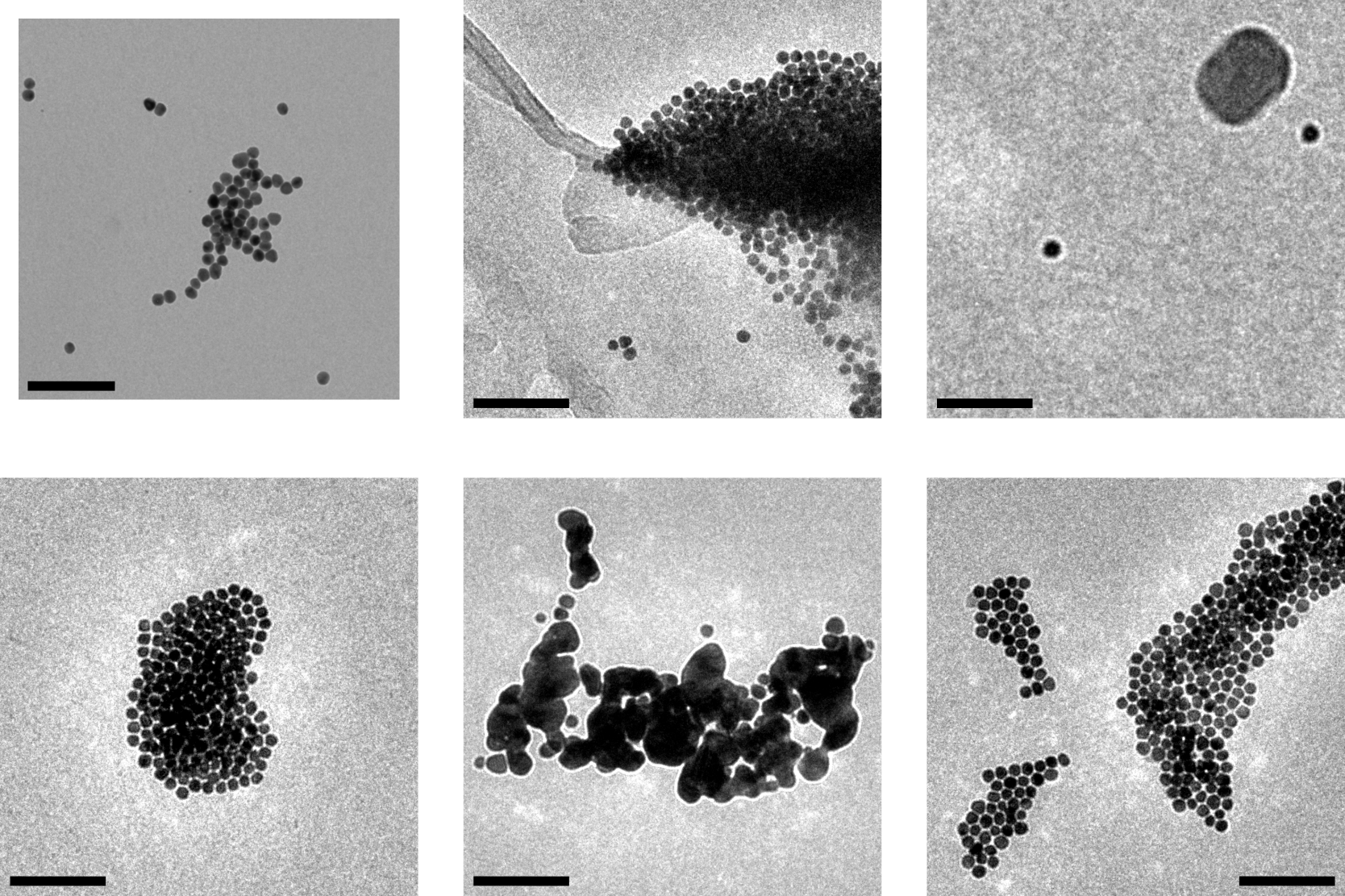}
	\caption{Aging time series of L-MPA after \SI{6}{h} containing various salts of \SI{50}{mM} concentration: (a) salt-free L-MPA suspension, (b--f) salts along the Hofmeister series as before (NaF, NaCl, NaBr, NaI, NaSCN). The scale bars in all images represent \SI{100}{nm}. Aggregation took place in suspension and aggregates assembled on a TEM grid for \SI{15}{min}. \newline 
		Especially for NaF (b) and NaCl (c) artifacts due to extended drying times are visible and the NaCl sample only displays two AuNPs (the rectangular feature is a drying artifact) as the adsorption of L-MPA to the TEM grids is much lower.}
	\label{si:TEM_LMPA}
\end{figure}

\begin{figure}
	\includegraphics[width=0.75\textwidth]{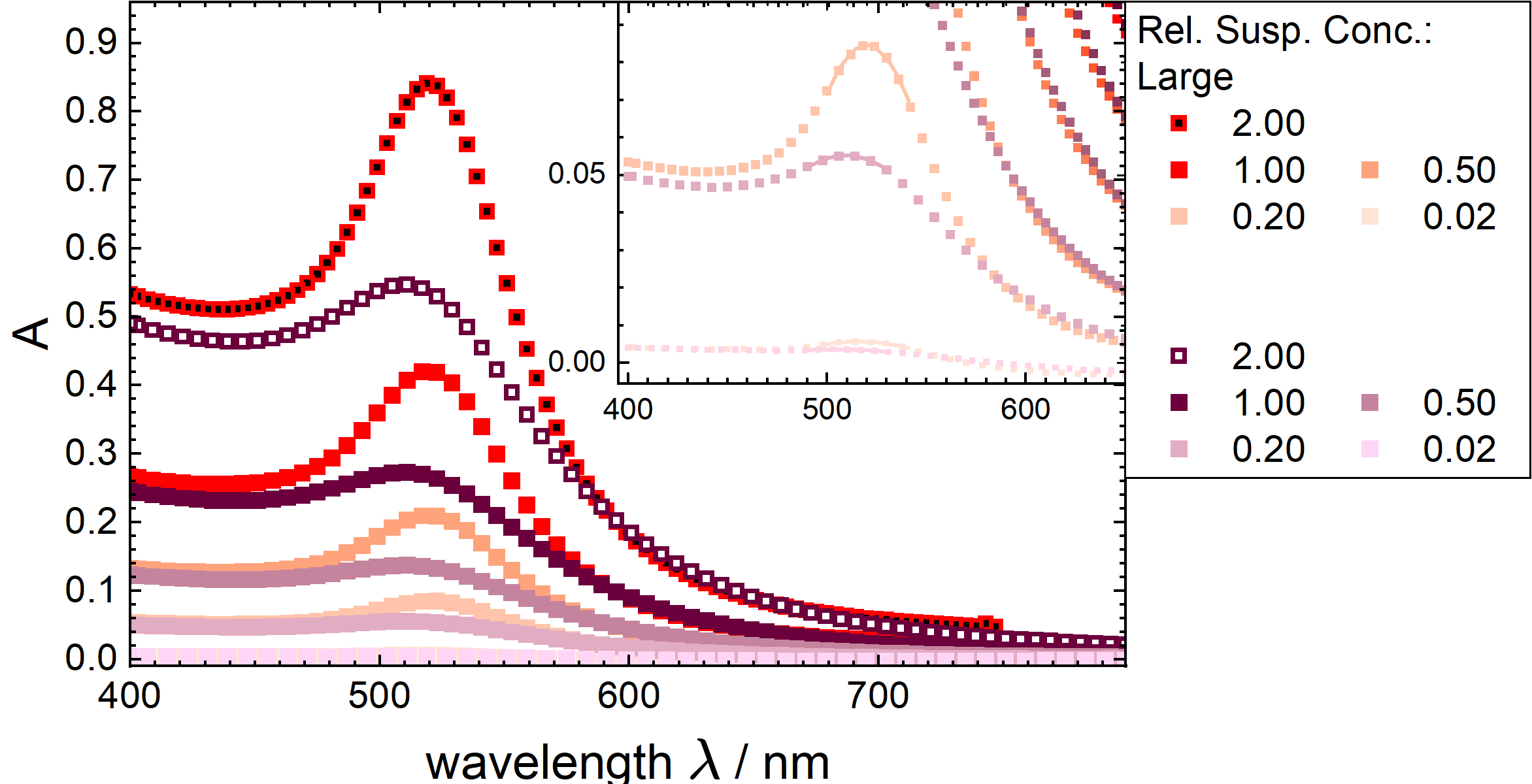}
	\caption{Absorption spectra of salt-free \mbox{L-Cit} and \mbox{S-Cit} suspensions at various AuNP concentrations. The AuNP concentration used in all other measurements is marked in the same color as the reference in those cases. \mbox{L-MPA} suspension series is not plotted here for clarity but behaves identical. }
	\label{si:AuNPDilutionSeries}
\end{figure}
\end{suppinfo}

\end{document}